\documentclass[iop]{emulateapj}
\usepackage{graphicx}
\usepackage{amssymb}
\usepackage{lscape}

\begin{document}

\title{$z\sim1$ Ly$\alpha$ Emitters I. The Luminosity Function\altaffilmark{1, 2, 3, 4}}

\author{Isak G. B. Wold\altaffilmark{5}, Amy J. Barger\altaffilmark{5,6,7},
and Lennox L. Cowie\altaffilmark{7}}

\altaffiltext{1}{Based in part on data obtained from the Mikulski Archive for Space Telescopes (MAST). STScI is operated by the Association of Universities for Research in Astronomy, Inc., under NASA contract NAS5-26555. Support for MAST for non-HST data is provided by the NASA Office of Space Science via grant NNX09AF08G and by other grants and contracts.}
\altaffiltext{2}{The W.~M.~Keck Observatory is operated as a scientific partnership among the the California Institute of Technology, the University of California, and NASA, and was made possible by the generous financial support of the W.~M.~Keck Foundation.}
\altaffiltext{3}{Some of the observations reported in this paper were obtained with the Southern African Large Telescope (SALT).}
\altaffiltext{4}{Based in part on zCOSMOS observations carried out using the Very Large Telescope at the ESO Paranal Observatory under Programme ID: LP175.A-0839.}
\altaffiltext{5}{Department of Astronomy, University of Wisconsin-Madison, 475 North Charter Street, Madison, WI 53706, USA; wold@astro.wisc.edu, barger@astro.wisc.edu}
\altaffiltext{6}{Department of Physics and Astronomy, University of Hawaii, 2505 Correa Road, Honolulu, HI 96822, USA}
\altaffiltext{7}{Institute for Astronomy, University of Hawaii, 2680 Woodlawn Drive, Honolulu, HI 96822, USA} 
\begin{abstract}
We construct a flux-limited sample of 135 candidate $z\sim1$ Ly$\alpha$
emitters (LAEs) from \textit{Galaxy Evolution Explorer} (\textit{GALEX})
grism data using a new data cube search method. These LAEs have luminosities
comparable to those at high redshifts and lie within a 7 Gyr gap present
in existing LAE samples. We use archival and newly obtained optical
spectra to verify the UV redshifts of these LAEs. We use the combination
of the \textit{GALEX} UV spectra, optical spectra, and X-ray imaging
data to estimate the active galactic nucleus (AGN) fraction and its
dependence on Ly$\alpha$ luminosity. We remove the AGNs and compute
the luminosity function (LF) from 60 $z\sim1$ LAE galaxies. We find
that the best fit LF implies a luminosity density increase by a factor
of $\sim$1.5 from $z\sim0.3$ to $z\sim1$ and $\sim20$ from $z\sim1$
to $z\sim2$. We find a $z\sim1$ volumetric Ly$\alpha$ escape fraction
of $0.7\pm0.4\%$.
\end{abstract}


\section{Introduction }

Ly$\alpha$ emitters (LAEs) have the potential to be a powerful cosmological
probe. The evolution of their number density at \textbf{$z\gtrsim7$
}offers the opportunity to constrain the opacity of the intergalactic
medium (IGM) and the timing of cosmic reionization \citep[e.g.,][]{stark11}.
Their clustering properties may be able to constrain efficiently the
expansion history of the universe \citep[e.g.,][]{hill08,greig13}.
Furthermore, at the highest redshifts ($z>6$) Ly$\alpha$ is the
only emission line that we can detect from the ground, making it the
only probe of the internal structure of these galaxies. However, to
use LAEs as probes of the high-redshift universe, we must first understand
their physical properties.

The Ly$\alpha$ line is resonantly scattered by neutral hydrogen,
making its flux and line profile notoriously hard to interpret \citep[e.g.,][]{neufeld91,kunth03,finkelstein07,schaerer08,ostlin09}.
For example, unlike rest-frame UV continuum observations, one cannot
merely apply a dust extinction correction to obtain an estimate of
the star formation rate (SFR). Thus, constraints on the Ly$\alpha$
escape fraction must be empirically measured. Furthermore, observational
studies are required to determine which physical properties facilitate
the escape of Ly$\alpha$ emission. 

Due to the limited resources in the UV, most observational studies
have focused on constraining the physical properties of $z\sim2-3$
LAEs, where the Ly$\alpha$ line moves into the optical. Detailed
analysis of the host galaxies is difficult at these high redshifts,
and different groups have come to drastically different conclusions
about them (e.g., compare \citealt[][]{hu98,nilsson07,gawiser07}
to \citealt{lai08,kornei10}). Furthermore, at these redshifts Ly$\alpha$
emission may be significantly altered by the intervening IGM. For
example, based on hydro-simulations, \citet[][]{laursen11} find that
only $\sim70\%$, $26\%$, and $20\%$ of Ly$\alpha$ photons are
transmitted through the IGM at $z\sim3.5$, $5.8$, and $6.5$, respectively.
To make progress, we need to study lower-redshift samples where LAEs
are bright due to smaller luminosity distances and where LAEs can
be integrated into comprehensive studies of galaxies at the same redshifts
to understand how LAEs are drawn from the general galaxy population. 

Recently, a $z$ $\sim$ 0.3 LAE sample has been found by searching
for emission-line objects in \textit{Galaxy Evolution Explorer} \citep[{\em GALEX\/};][]{martin05}
FUV pipeline spectra. Since the \textit{GALEX} grism pipeline only
includes objects whose UV continuum magnitudes are bright enough to
generate measurable continuum spectra, only a fraction of the emission-line
objects are extracted. Thus, this sample's selection process is most
analogous to locating LAEs in the high-redshift Lyman break galaxy
(LBG) population via spectroscopy \citep[e.g.,][]{shapley03}. However,
the procedure enables the selection of a substantial sample of $z$
$\sim$ 0.3 sources, and many papers have investigated their properties
\citep[e.g.,][]{deharveng08,finkelstein09,atek09,scarlata09,cowie10,cowie11}.
The picture that emerged from the optical follow-up of the $z\sim0.3$
\textit{GALEX} sample is that LAEs, when compared to UV-continuum
selected galaxies, are relatively young, compact, metal poor, star-forming
galaxies \citep[][]{cowie10,cowie11}.

However, it is not clear that the properties of these low-redshift
($z\sim0.3$) sources are representative of high-redshift LAEs. In
particular, LAEs are considerably fainter and much less common at
$z\sim0.3$ than they were in the past, with only about $5\%$ of
$z\sim0.3$ UV-continuum selected galaxies having rest-frame Ly$\alpha$
equivalent widths (EW$_{{\rm {r}}}$(Ly$\alpha$)) greater than $20$~\AA,
compared with $20\%-25\%$ at $z\sim3$ \citep[][]{shapley03}. The
EW$_{{\rm {r}}}$(Ly$\alpha$) $\geq$ 20\AA~ requirement is typically
used to define high-redshift samples \citep[e.g.,][]{hu98}. The
first redshift where LAEs are seen that are comparable in luminosity
to high-redshift LAEs is at $z$ $\sim$ 1. However, only a handful
of $z\sim1$ LAEs have \textit{GALEX }detectable continuum spectra.
Thus, we need to remove the pipeline's continuum selection requirement
(NUV$<$22) in order to study a large sample of $z$ $\sim$ 1 LAEs.
This has the additional virtue of producing a purely flux-limited
sample that is straightforward to analyze. \begin{deluxetable*}{ccccccc} 
\tablecolumns{7} 
\tablewidth{0pc} 

\tablecaption{\textit{GALEX} field exposure time, $80\%$ completeness flux threshold, number of LAE candidates, and number of confirmed LAE galaxies} 
\tablehead{ \colhead{\textit{GALEX} Field} & \colhead{$\alpha$(J2000.0)} & \colhead{$\delta$(J2000.0)} & \colhead{Exp. time} & \colhead{$f_{\rm{Ly}\alpha}$ (erg cm$^{-2}$ s$^{-1}$)} & \colhead{Num. LAE candidates}& \colhead{Num. confirmed LAEs}\\ 
\colhead{(1)} & \colhead{(2)} & \colhead{(3)} & \colhead{(4)} & \colhead{(5)} & \colhead{(6)}& \colhead{(7)}} 
\startdata 
CDFS & 3$^h$30$^m$40$^s$ & -27$^\circ$27$'$43$''$ & 353 ks & 1.8$\times$10$^{-15}$ & 40 & 24\\ 
GROTH & 14$^{h}$19$^{m}$58$^{s}$ & 52$^{\circ}$46$'$54$''$ & 291 ks & 2.0$\times$10$^{-15}$ & 41 & 19\\ 
NGPDWS & 14$^{h}$36$^{m}$37$^{s}$ & 35$^{\circ}$10$'$17$''$ & 165 ks & 2.3$\times$10$^{-15}$ & 35 & 11\\ 
COSMOS & 10$^{h}$00$^{m}$29$^{s}$& +2$^{\circ}$12$'$21$''$ & 140 ks & 3.2$\times$10$^{-15}$ & 19 & 4\\ 
\enddata 
\label{fields}
\end{deluxetable*}

In \citet*[][hereafter, BCW12]{barger12}, we solved the problem of
obtaining a flux-limited sample of $z\sim1$ LAEs for the \textit{Chandra}
Deep Field-South (CDFS). To do this, we converted the multiple \textit{GALEX}
grism images into a three-dimensional (two spatial axes and one wavelength
axis) data cube. We then treated the wavelength ``slices'' as narrowband
images in which we searched for emission-line galaxies. Through simulations,
we showed that we could recover more than 80\% of the sources with
$f$(Ly$\alpha$) $\gtrsim$ 10$^{-15}$ erg cm$^{-2}$ s$^{-1}$,
which corresponds to $L$(Ly$\alpha$) $\gtrsim$ 10$^{42.5}$erg
s$^{-1}$ at $z\sim1$. By comparing our sample to X-ray data, existing
optical spectroscopy, and deep {\em U\/}-band imaging, we determined
that nearly all of our 28 new LAEs are real and that the
UV spectroscopic redshifts based on the Ly$\alpha$ identifications
are reliable. We also determined the fraction of active galactic nuclei
(AGNs) in the sample relative to star formers.

In this paper, we apply the data cube search method to all of the
deepest \textit{GALEX} grism fields (CDFS-00, 353 ks; GROTH-00, 291
ks; NGPDWS-00, 165 ks; COSMOS-00, 140 ks), which correspond to some
of the most intensively studied regions in the sky. This work provides
the first large sample of $z=0.67-1.16$ LAEs (N=60) that can be used
to investigate the physical properties of these galaxies. There are
relatively large samples of known LAEs at redshifts of $z\sim0.3$
and $z\sim2$. Thus, a sample of $z\sim1$ LAEs is needed to map the
evolution of LAEs over a $\sim$7 Gyr gap. This redshift regime is
where the star-forming properties of galaxies change very rapidly
and where the star formation begins to decline. It is a key area for
connecting to the high-redshift universe. 

In Paper II, we will leverage existing archival spectra and followup
optical spectra to constrain their physical properties. In particular,
we will study which properties facilitate the escape of Ly$\alpha$
emission and how LAEs are drawn from the overall galaxy population.
This will help us understand how the Ly$\alpha$ emission properties
of galaxies evolve as we move to higher redshifts and higher Ly$\alpha$
luminosities.

Here we catalog our candidate $z\sim1$ LAE samples in each field
and give optical redshifts from both archival and newly obtained observations.
With X-ray, UV, and optical data, we determine the false detection
rate (cases where the emission line is either not confirmed or is
not Ly$\alpha$) and the AGN contamination rate of our sample. With
the remaining LAEs, we compute the LAE galaxy luminosity function
(LF) at $z\sim1$ and use this to investigate the evolution of the
Ly$\alpha$ LF and the Ly$\alpha$ escape fraction over the redshift
range from $z\sim0.3$ to $z\sim2$. Unless otherwise noted, we give
all magnitudes in the AB magnitude system ($m_{\mbox{AB}}=23.9-2.5$log$_{10}f_{\nu}$
with $f_{\nu}$ in units of $\mu$Jy). We use a standard $H_{0}=70$
km s$^{-1}$ Mpc$^{-1}$, $\Omega_{M}=0.3$, and $\Omega_{\Lambda}=0.7$
cosmology.

\section{Choice of fields}

 We study the four deepest NUV grism observations: CDFS-00, GROTH-00,
NGPDWS-00, and COSMOS\nolinebreak[4]-\nolinebreak[4]\nopagebreak[4]00.
The {\em GALEX\/} fields are large ($\sim1$~deg$^2$), but in subregions
of the fields, there are many objects with optical spectra (CDFS,
\citealt{vanzella08}, \citealt{popesso09}; GROTH, \citealt{newman12};
COSMOS, \citealt{lilly07}, \citealt{trump09}). In subregions, there
are also multi-color observations from the {\em HST\/} GOODS \citep{giavalisco04},
CANDELS \citep{grogin11,koekemoer11}, COSMOS \citep{leauthaud07},
and GEMS \citep{rix04} programs that provide the galaxy morphologies
and spectral energy distributions from the rest-frame far-UV to the
mid-infrared. When corrected for the emission-line contributions using
the spectra \citep[][]{schaerer09, atek11, cowie11}, the spectral
energy densities can be used to compute the ages and extinctions of
the galaxies. 

Finally, the 4~Ms {\em Chandra\/} image of the CDFS \citep[][]{alexander03, luo08}
region, along with shallower X-ray observations in the Extended CDFS
\citep[][]{lehmer05, virani06}, COSMOS \citep[][]{elvis09}, GROTH
\citep[][]{laird09}, and NGPDWS \citep[][]{kenter05} fields, can
be used to identify AGNs.  AGNs may also be identified using the
UV grism spectra obtained with \textit{GALEX} and the optical spectra,
but the X-rays provide a valuable cross-check, where they are available.

\section{\textit{GALEX} NUV LAEs}

\begin{figure*}
\includegraphics[bb=90bp 75bp 710bp 525bp,clip,angle=180,scale=0.4]{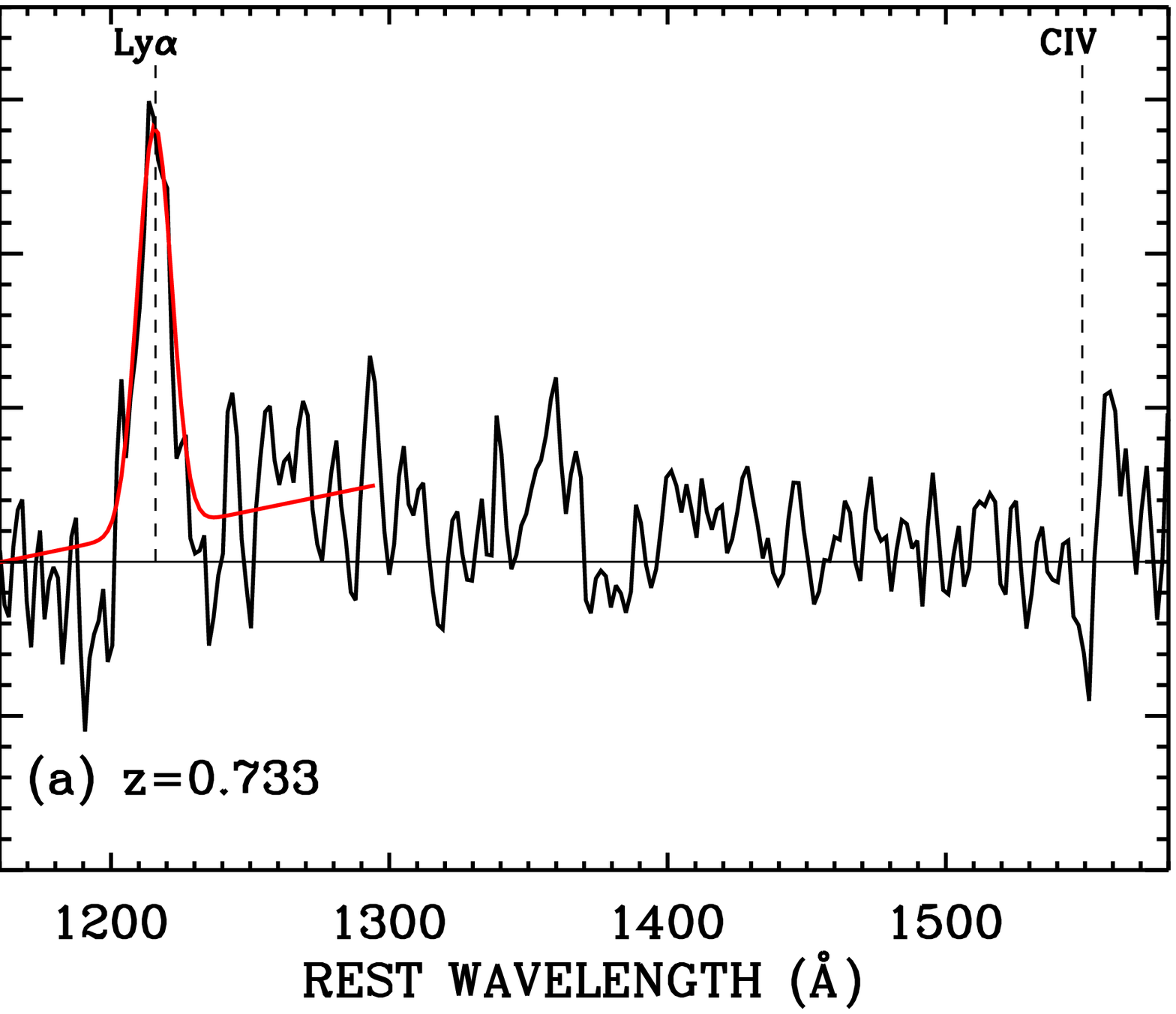}\includegraphics[bb=90bp 75bp 680bp 525bp,clip,angle=180,scale=0.4]{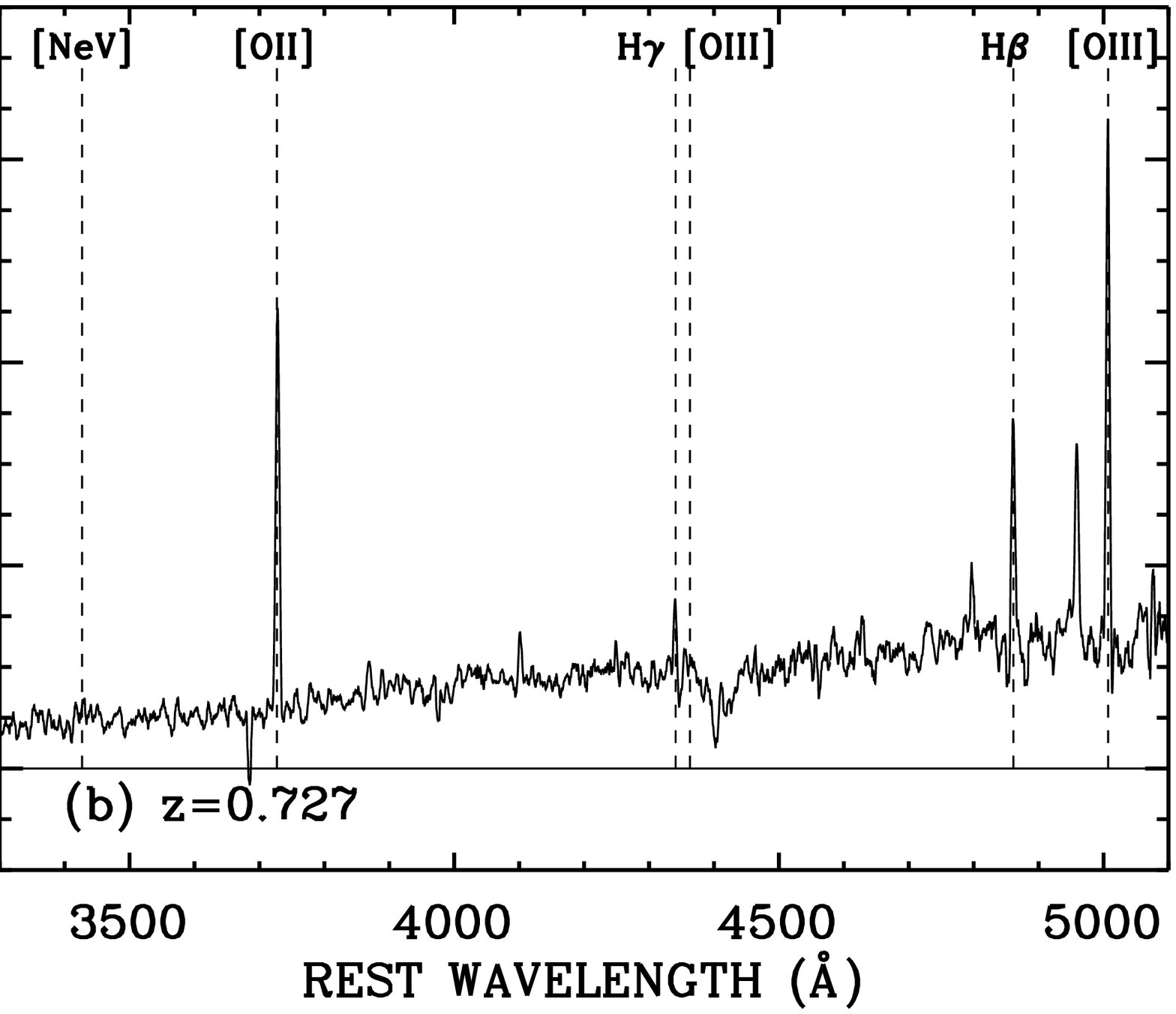}\caption{\textbf{(a)} LAE galaxy UV spectrum GALEX033100-273020. The redshift
($z=0.733$) and Ly$\alpha$ flux is measured from the Gaussian fit
(red profile). The lack of detected C{\small IV}$\lambda1549$ is
consistent with this source being star-forming galaxy rather than
AGN. \textbf{(b)} Keck-DEIMOS spectrum for the same source. The optical
redshift ($z=0.727$) is found to be consistent with the UV redshift.
The lack of detected {[}Ne{\small V}{]}$\lambda3426$ or broad Balmer
lines is consistent with this source being star-forming galaxy rather
than AGN. }

\label{ex}

{\footnotesize (A color version of this figure is available in the online journal.)}
\end{figure*}

\subsection{Catalog Extraction}

As described in BCW12, we converted multiple \textit{\small GALEX}
NUV grism images into a three-dimensional (two spatial axes and one
wavelength axis) data cube. Each background subtracted data cube consists
of thirty $20$ \AA\ narrowband slices covering a wavelength range
of 2030 to 2630 \AA\ and a 50$'$$\times$50$'$ field of view.
We used SExtractor \citep{bertin96} to identify all 4$\sigma$ sources
within the cube and then visually inspected each source and its spectrum
(1-D and 2-D) to eliminate objects that were artifacts. Applying the
data cube search produced 40 CDFS, 41 GROTH, 35 NGPDWS, and 19 COSMOS
objects for a full Ly$\alpha$ selected candidate sample of 135 (see
Table \ref{fields}). In Figure \ref{ex}(a), we show the extracted
1-D spectrum for GALEX033100-273020 to illustrate the quality of our
\textit{GALEX} spectra. BCW12 performed this routine on the deepest
NUV grism field (CDFS-00). To verify the BCW12 sample of 28 LAE candidates,
we have independently searched the CDFS cube for emission-line sources.
We found 27 of the 28 objects that were identified by BCW12, and we
found 12 new objects.  Approximately half of these new LAE candidates
are close to the field's $80\%$ completeness flux threshold (see
Section 3.3) and were probably missed due to their faintness. The
other half are relatively bright and were probably mis-classified
as data cube artifacts, such as the remaining edge effects from brighter
objects. The BCW12 LAE candidate missed by this search (GALEX033124-275625)
is real and has a consistent optical redshift. This LAE was eliminated
in the visual inspection phase of our search due to its proximity
to a UV-bright star. We include this missed LAE in all subsequent
analysis. To date, we have confirmed the UV redshifts of 92$\%$ (11
out of 12) of the newly discovered CDFS LAEs with optical follow-up
spectra.

\subsection{Optical Spectroscopic Followup}

Spectroscopic observations were primarily obtained with the DEep Imaging
Multi-Object Spectrograph \citep[DEIMOS;][]{faber03} on Keck II in
a series of runs in 2012 to 2013. The observations were made with
the ZD600 line mm$^{-1}$ grating blazed at 7500 \AA. This gives
a resolution of $\sim$ 5 \AA~ with a 1$''$ slit and a wavelength
coverage of 5300 \AA. The high spectral resolution is necessary to
distinguish the {[}O{\small II}{]}$\lambda$3727 doublet structure.
The observations were not generally taken at parallactic angle, since
the position angle was determined by the mask orientation. Each $\sim$30
min exposure was broken into three subsets, with the objects stepped
along the slit by 1.5$''$ in each direction. The raw two-dimensional
spectra were reduced and extracted using the procedure described in
\citet{cowie96}. In Figure \ref{ex}(b), we show the optical spectrum
for GALEX033100-273020 to illustrate the quality of our\textit{ }DEIMOS
spectra. We used the remaining space available in our MOS masks to
observe a control sample, a UV-continuum selected sample without detected
Ly$\alpha$ that have the same luminosities and are expected, based
on their colors, to lie in the same redshift interval as our primary
LAE sample. In Paper II, we will use the control sample to determine
how the LAE galaxies are drawn from the general galaxy population. 

\begin{figure}
\includegraphics[bb=122bp 75bp 690bp 525bp,clip,angle=180,scale=0.4]{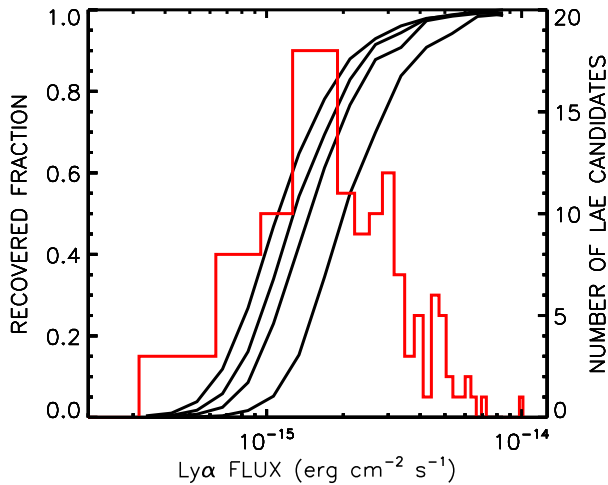}\caption{Fraction of simulated sources recovered as a function of the emission-line
flux. From left to right the black curves show recovered fractions
from CDFS, GROTH, NGPDWS, and COSMOS fields (also see Table \ref{fields}).
The red histogram shows the number of LAE candidates as a function
of flux for all four \textit{GALEX} fields.}

\label{compl}

{\footnotesize (A color version of this figure is available in the online journal.)}
\end{figure}

We observed our southern and equatorial fields (COSMOS and CDFS) with
the Robert Stobie Spectrograph \citep[RSS;][]{burgh03} on SALT from
2011 to 2013. The observations were made with the pg1300 grating with
the grating angle adjusted to ensure spectral coverage of lines redward
of {[}NeV{]}. This gives a resolution of $\sim$ 4 \AA~ with a 1.5$''$
slit and a wavelength coverage of 2000 \AA. We observed each target
for $\sim$30 min. The raw two-dimensional spectra were reduced and
extracted using the IRAF packages LONGSLIT and APEXTRACT.

\subsection{Catalog Completeness}

To determine the limitations of our multi-field catalogs and to compute
the LAE galaxy LF, we have developed a simulation to determine the
completeness of recovery versus flux. For each field, we added 1000
simulated emitters uniformly within the field's data cube. We did
not model morphology or size difference, since nearly all emitters
are unresolved at the spatial ($\sim6''$) and spectral resolution
($\sim25$ \AA) of the \textit{\small GALEX} grism data. We then
ran our standard selection procedure and found the number of recovered
objects. We independently performed the above procedure ten times,
giving a total of 10,000 input sources. In Table \ref{fields}, we
list the flux threshold above which each field is greater than 80$\%$
complete. As expected, the completeness limit scales as the inverse
square root of the exposure time. Below this threshold, the completeness
of our sample rapidly declines (see Figure \ref{compl}). 
\begin{figure}
\includegraphics[bb=70bp 75bp 700bp 530bp,clip,angle=180,scale=0.4]{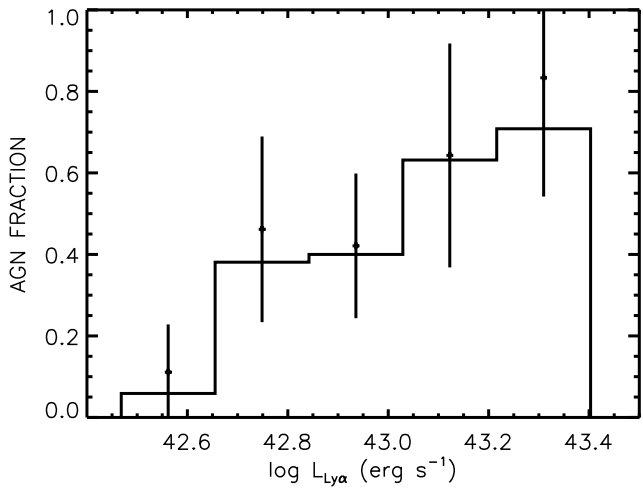}\caption{AGN fraction per Ly$\alpha$ luminosity bin for the EW$_{{\rm r}}$(Ly$\alpha$)
$\geq$ 20 \AA~ LAE data cube sample. The histogram shows the sample
wide AGN fractions, while the points with error bars show the AGN
fractions for regions with deep X-ray data.}

\label{fagn}
\end{figure}
\begin{figure*}
\includegraphics[bb=70bp 75bp 700bp 520bp,clip,angle=180,scale=0.4]{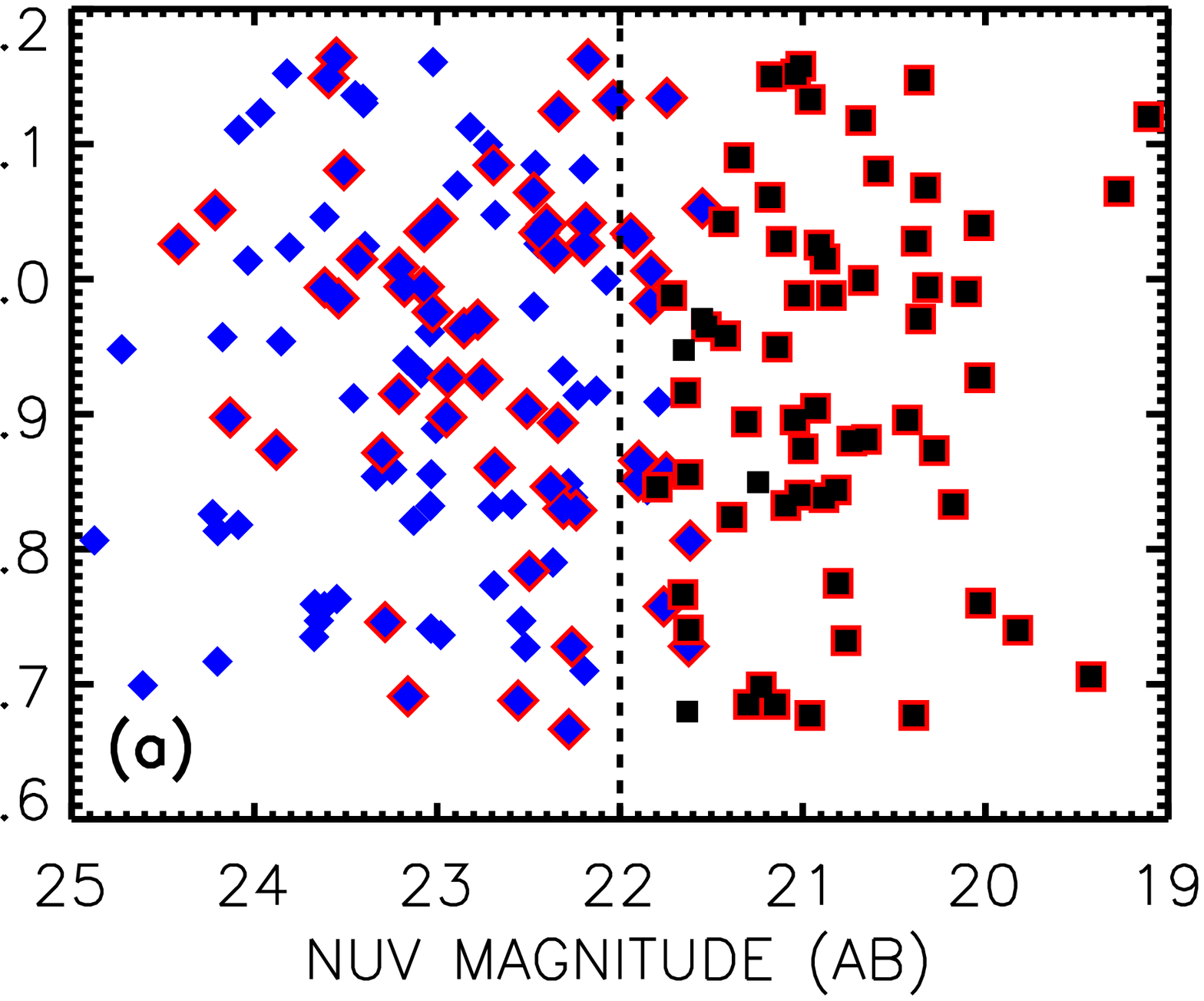}\includegraphics[bb=70bp 75bp 700bp 525bp,clip,angle=180,scale=0.4]{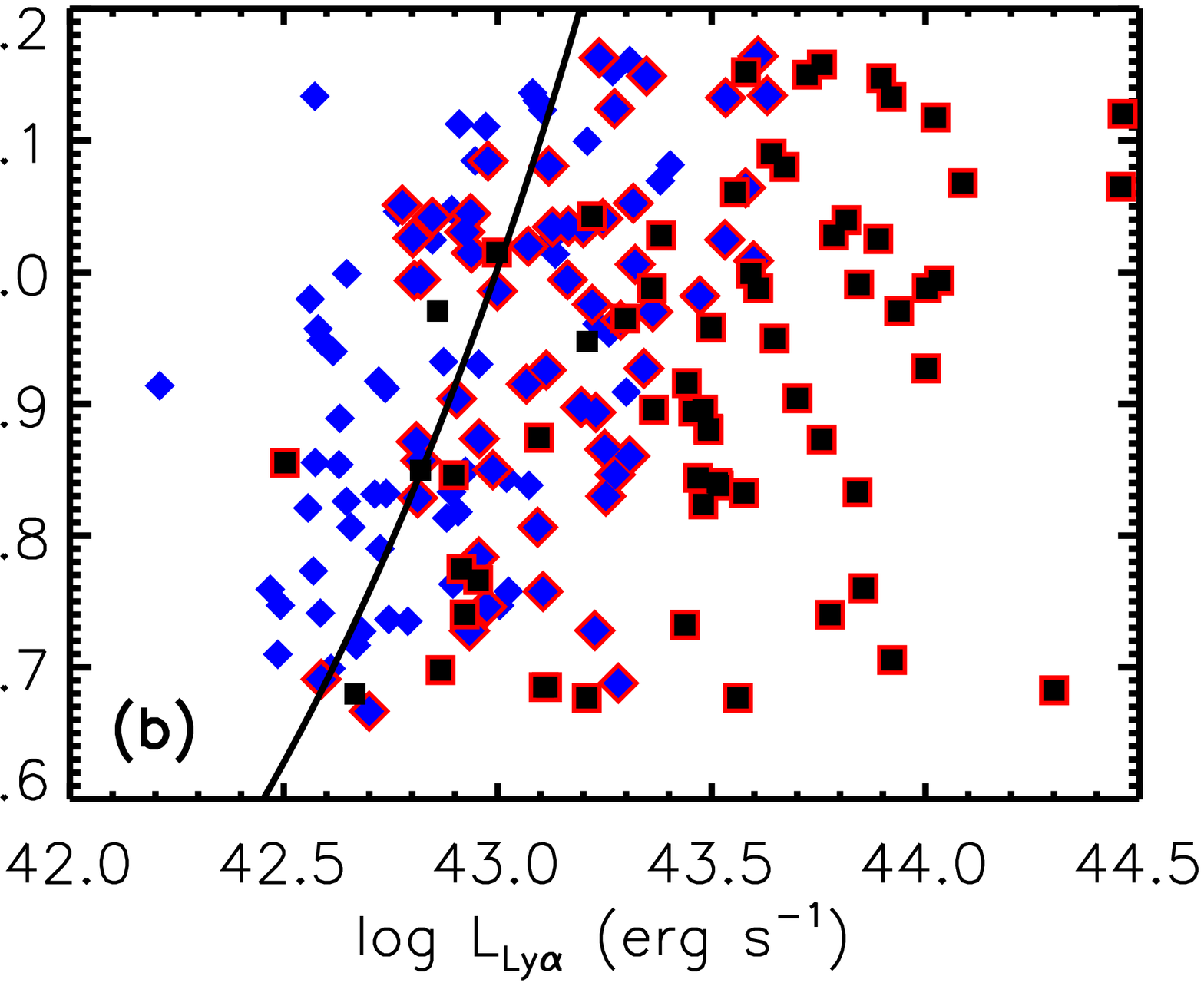}

\includegraphics[bb=70bp 75bp 700bp 520bp,clip,angle=180,scale=0.4]{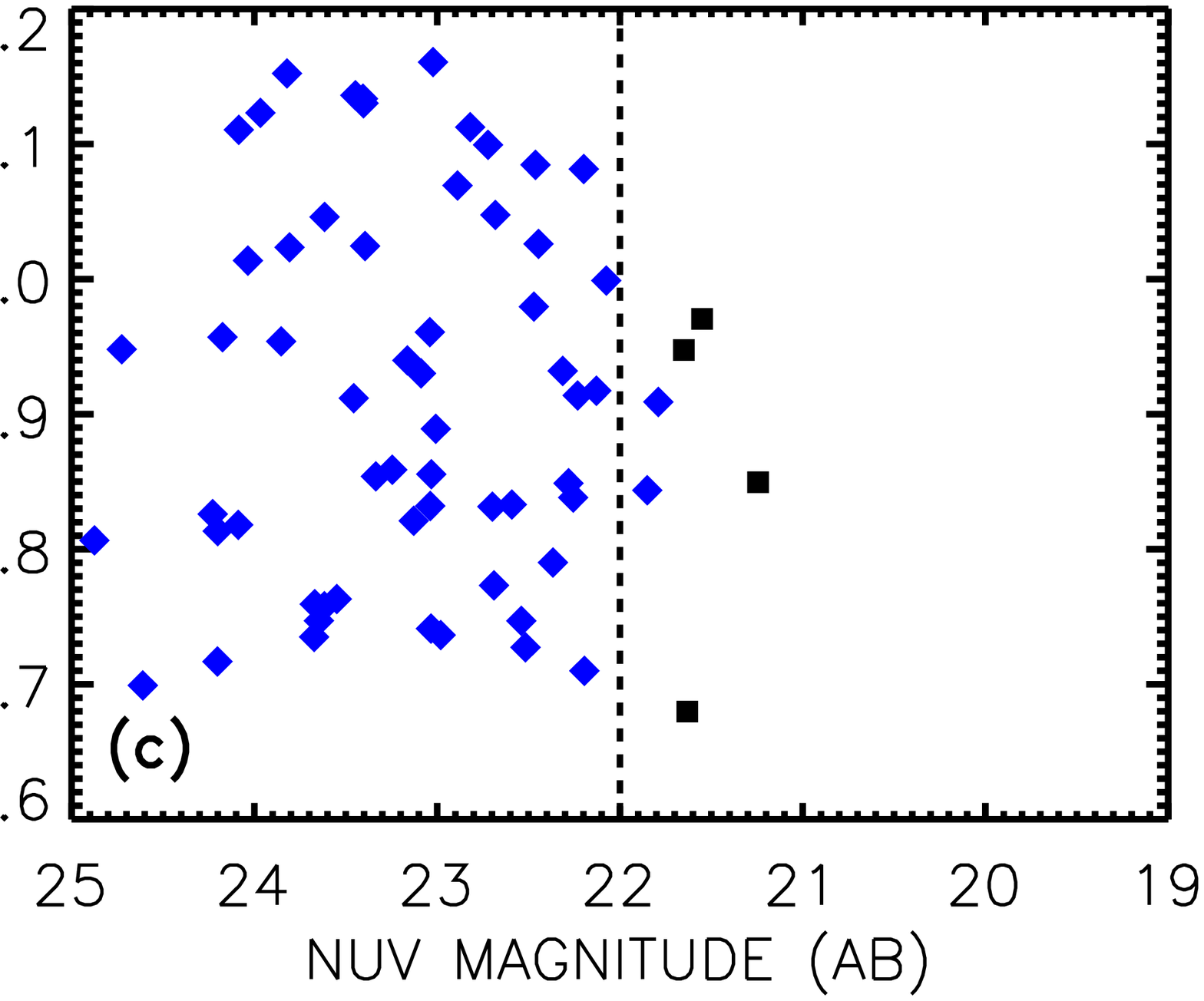}\includegraphics[bb=70bp 75bp 700bp 525bp,clip,angle=180,scale=0.4]{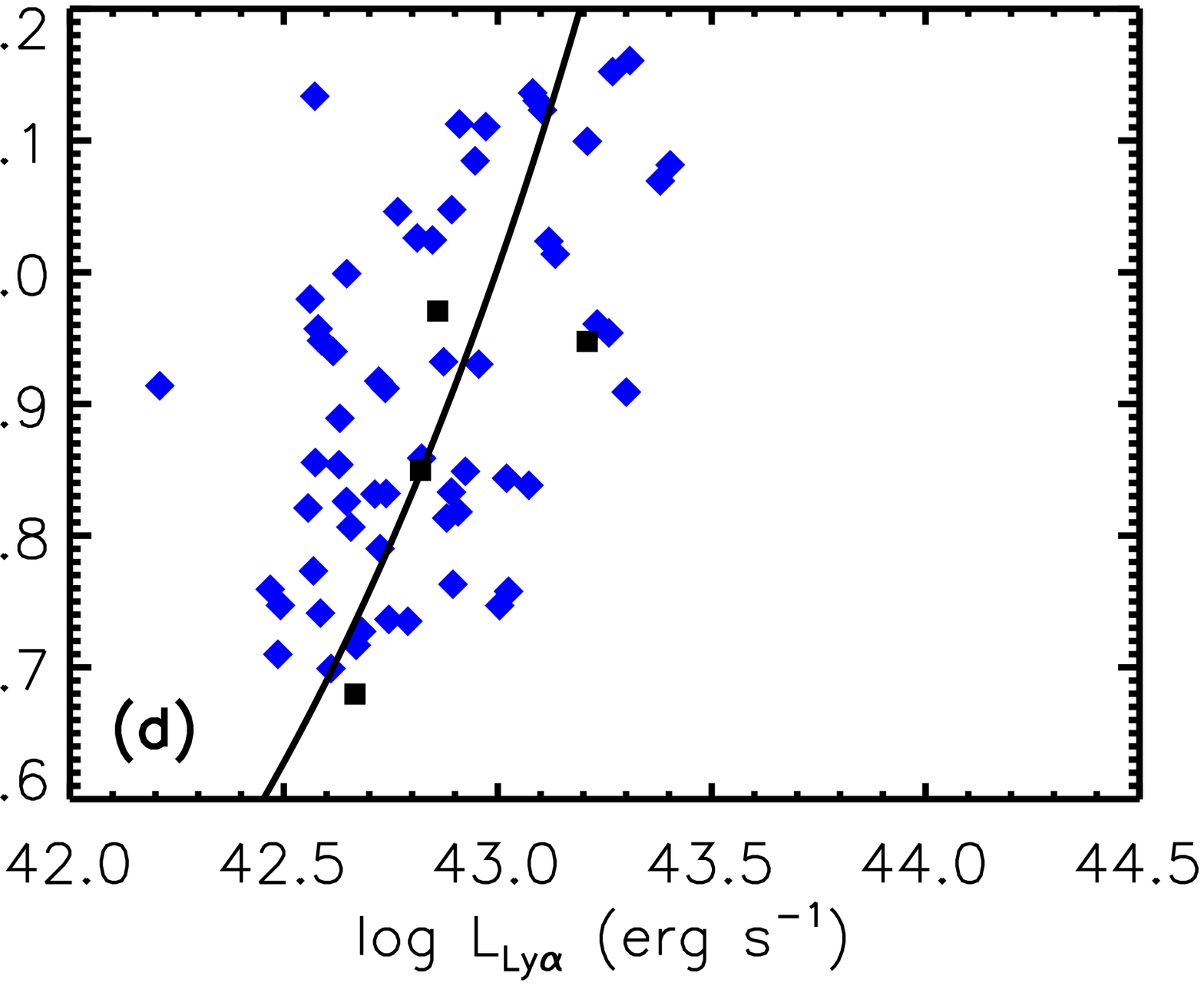}

\caption{\textbf{(a)} LAE redshift vs.\ Ly$\alpha$ luminosity. Black squares
show the pipeline LAEs found by \citet{cowie10,cowie11} constrained
to the data cubes' 50$'$$\times$50$'$ FOVs and redshift range ($z=0.67-1.16$).
Blue diamonds show our data cube LAEs. Objects classified as AGNs
in any way are surrounded by red. The dashed line shows the \textit{GALEX}
pipeline's magnitude limit NUV$\sim22$.\textbf{\textit{ }}\textbf{(b)
}LAE redshift vs. Ly$\alpha$ luminosity. The solid line indicates
the data cube's $\sim$80\% flux completeness limit. \textbf{(c) }The
same as Figure \ref{pipe}(a), but with all AGNs removed. \textbf{(d)}
The same as Figure \ref{pipe}(b), but with all AGNs removed.}

\label{pipe}

{\footnotesize (A color version of this figure is available in the online journal.)}
\end{figure*}

\subsection{Catalogs of LAE Candidates by Field}

In Tables \ref{cdfs}-\ref{cosmos}, we list all of the LAE candidates
in the CDFS, GROTH, NGPDWS, and COSMOS fields ordered by redshift.
In Column 1, we give the \textit{GALEX} name; in Columns 2 and 3,
the J2000 right ascension and declination based on NUV position; in
Columns 4 and 5, the NUV and FUV AB magnitudes; in Column 6, the redshift
from the \textit{GALEX} UV spectrum; in Column 7, the logarithm of
the Ly$\alpha$ luminosity; in Column 8, the rest-frame EW$_{{\rm r}}$(Ly$\alpha$)
with 1$\sigma$ errors; in Column 9, the logarithm of the X-ray flux;
in Column 10 , the UV classified AGNs; in Columns 11 and 12, the J2000
right ascension and declination based on optical position; in Column
13, the offset between the optical to UV positions; in Column 14,
the optical redshift; and in Column 15 the optically classified AGNs.

We measured NUV and FUV AB magnitudes from the archival \textit{GALEX}
background subtracted intensity maps \citep[][]{morrissey07}. We
first determined the magnitudes within 8$''$ diameter apertures centered
on each of the emitter positions. To correct for flux that falls outside
our apertures, we measured the offset between 8$''$ aperture magnitudes
and \textit{GALEX} pipeline total magnitudes for all bright cataloged
objects (20-23 mag range) within our fields. We determined the median
offset for each field and applied these to the magnitudes listed in
Columns 4 and 5. All objects are bright in the NUV when compared to
their FUV magnitudes. In some cases, we measure negative FUV fluxes.
For these cases, we list the magnitude corresponding to the absolute
value of the flux with a minus sign in front to indicate that the
flux was negative.

We corrected our one-dimensional NUV spectra for Galactic extinction
assuming a \citet[][]{fitzpatrick99} reddening law with R$_{V}$=3.1.
We obtained $A_{V}$ values from the \citet[][]{schlafly11} recalibration
of the \citet[][]{cardelli89} extinction map as listed in the NASA/IPAC
Extragalactic Database (NED). Galactic extinction modifies the Ly$\alpha$
flux by $\sim$11\% for the COSMOS LAEs, $\sim$4\% for the GROTH
LAEs, and $\sim$6\% for the CDFS and NGPDWS LAEs.

From these extinction corrected spectra, we measured the redshifts,
the Ly$\alpha$ fluxes, and the line widths using a two step process.
First, we fit a 140 \AA~ rest-frame region around the Ly$\alpha$
line with a Gaussian and a sloped continuum (see Figure \ref{ex}(a)).
A downhill simplex optimization routine was used to $\chi^{2}$ fit
the five free parameters (continuum level and slope plus Gaussian
center, width, and area). We used the results of this fitting process
to eliminate the two continuum parameters and as a starting point
for the second step. In the second step, we used the IDL MPFIT procedures
of \citet[][]{markwardt09} to $\chi^{2}$ fit the remaining three
Gaussian parameters. We also employed the same fitting routine but
with only a flat continuum (one continuum free parameter instead of
two). We find that our results are not significantly affected by this
model assumption. With the best-fit redshifts and Ly$\alpha$ fluxes,
we calculated Ly$\alpha$ luminosities. When available, we used the
more precise optical redshift rather than the UV redshift to calculate
the Ly$\alpha$ luminosities. We list the UV redshifts and Ly$\alpha$
luminosities in Columns 6 and 7.

The rest-frame EW$_{{\rm r}}$(Ly$\alpha$) measured on the spectra
are quite uncertain due to the very faint UV continuum. We obtained
a more accurate rest-frame EW by dividing the measured Ly$\alpha$
flux by the continuum flux measured from the broadband NUV image.
It is these rest-frame EWs that are given in Column 8. In Section
5, we investigate the $z\sim1$ rest-frame EW distribution. 

We made a classification of whether the emitter was an AGN based on
the presence of either broad or high-excitation emission lines in
its UV or optical spectra or on the presence of an X-ray counterpart.
Candidate X-ray counterparts were identified by matching all X-ray
sources within a $6''$ radius from the data cube position. We then
manually inspected the matches to reject false counterparts. We list
the X-ray flux of each identified counterpart in Column 9. The X-ray
luminosity threshold of $10^{42}$ erg s$^{-1}$ that is usually used
to define AGN activity \citep{hornschemeier01,barger02,szokoly04}
corresponds to an X-ray flux of $\gtrsim$10$^{-15.3}$ erg cm$^{-2}$
s$^{-1}$ for our $z>0.67$ sample. All X-ray detected sources exceed
this luminosity threshold, and we hereafter consider these sources
to be AGNs. In Column 10, we list sources classified as AGNs based
on the presence of broad or high-excitation emission lines in their
UV spectrum (see, e.g., \citealt{cowie10}). Our candidate star-forming
galaxy sample will still contain some remaining AGNs, so we used the
optical spectra to make a final determination of whether a galaxy
is, in fact, star-forming. In Column 15, we identify optical AGNs
based on the presence of either detectable {[}NeV{]} or broad Balmer
or Mg{\small II}$\lambda$2798 emission lines in the optical spectra.

We obtained optical redshifts for 122 of our 135 candidate LAEs. We
found that 15 optical redshifts did not match the measured UV redshifts.
For these objects, the Ly$\alpha$ luminosity and the rest-frame EW$_{{\rm r}}$(Ly$\alpha$)
fields are left blank in Columns 7 and 8. These sources are either
spurious, stars, or strong C{\small IV}$\lambda$1549 emitters. For
example, we found three LAE candidates in our COSMOS-00 field to be
$z\sim0.35$ C{\small IV}$\lambda$1549 emitters. We indicate these
objects by showing their optical redshift in parentheses or by setting
their optical redshift and type to `star' in Columns 14 and 15. Additionally,
we observed 4 candidate LAEs with Keck DEIMOS without recovering an
optical redshift. We indicate these objects by setting their optical
redshifts to `no z' in Column 14. For the purpose of computing the
LF, we retain the NUV brightest `no z' source (GALEX033045-274506),
which has two nearby untargeted potential optical counterparts. The
remaining `no z' candidate LAEs have relatively faint NUV counterparts
and may be spurious. Given the spatial resolution of our data cubes
($\sim6''$), we find that a total of five `no z' and non-matching
redshift sources have potential alternative optical counterparts.
We list these `{[}alt{]}' sources in the Table \ref{cdfs}-\ref{cosmos}
notes. Overall, we exclude 18 optically unconfirmed LAE candidates
in the following discussion and in the construction of the $z\sim1$
Ly$\alpha$ LF.
\begin{figure*}
\textbf{\textit{\includegraphics[bb=75bp 75bp 700bp 520bp,clip,angle=180,scale=0.4]{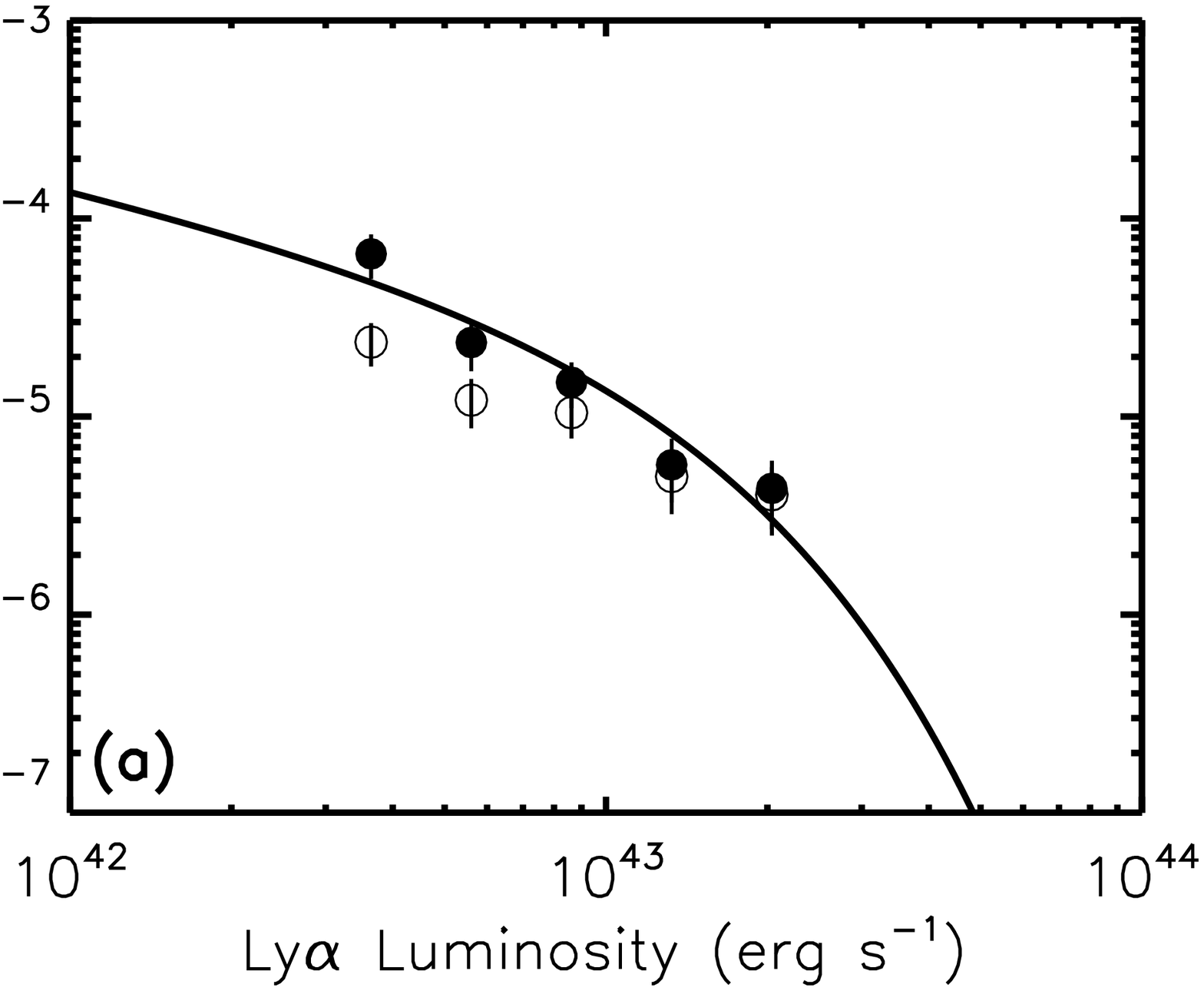}\includegraphics[bb=75bp 75bp 700bp 520bp,clip,angle=180,scale=0.4]{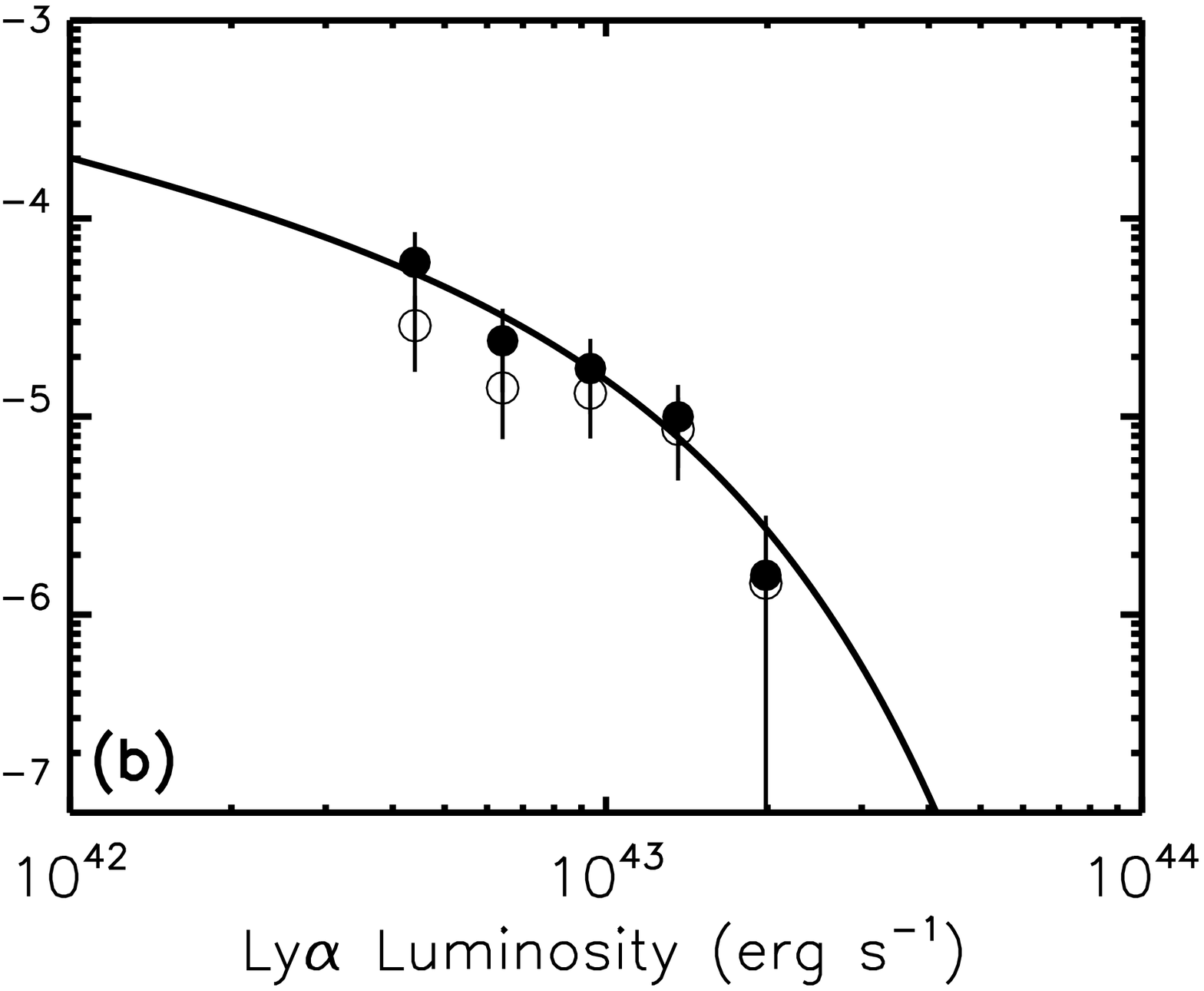}}}\caption{\textbf{(a)} Derived Ly$\alpha$ luminosity function at $z=0.67-1.16$
for the LAE galaxies in deep \textit{GALEX} grism fields (CDFS, GROTH,
NGPDWS, and COSMOS) with EW$_{{\rm {r}}}$(Ly$\alpha$)$\geq$20 \AA~
from both the \textit{GALEX} pipeline and the data cube samples (black
circles: open---raw data; solid---corrected for the effects of incompleteness
using the results from our Monte Carlo simulations). The black curve
indicates the best fit Schechter function assuming a fixed slope of
$\alpha=-1.6$. We find best fit parameters log $L_{\star}=43.0\pm0.2$
and log $\phi_{\star}=-4.8\pm0.3$. \textbf{(b)} Same as Figure \ref{lf}(a),
but with the LAE survey limited to regions with deep X-ray data to
ensure a robust AGN classification. Assuming a fixed slope of $\alpha=-1.6$,
we find best fit Schechter parameters log $L_{\star}=42.8\pm0.2$
and log $\phi_{\star}=-4.5\pm0.3$.}

\label{lf}
\end{figure*}

Excluding these 18, we are left with a sample of 117 LAEs. We classified
57 of these sources as UV, X-ray, or optical AGN. This establishes
an AGN fraction ($f$$_{{\rm {AGN}}}$) of 49$\%\pm$8$\%$ for our
sample. After we exclude UV identified AGNs, our $f$$_{{\rm {AGN}}}$
drops to 41$\%\pm$8$\%$. When performing the optical spectroscopic
followup, we preferentially targeted LAEs that were not classified
as UV or X-ray AGNs. With this policy, we obtained optical redshifts
for 98\% (59 out of 60) of our LAE galaxies but only 89$\%$ (51 out
of 57) of our AGNs. We found that 9 of the 37 LAEs classified as X-ray
AGNs were already classified as AGNs based on their UV spectra. We
observed 19 LAEs classified as X-ray AGNs with Keck DEIMOS. Only 3
of these X-ray AGNs were not classified as optical AGNs.

As a cross check to our sample's AGN fraction of 49$\%\pm$8$\%$,
we only consider LAEs within deep X-ray fields and re-compute the
AGN fraction. There are 79 candidate LAEs within deep X-ray fields;
we classified 44 (or 56$\%\pm$10$\%$) as AGNs. This is not significantly
different our sample-wide AGN contamination. In Figure \ref{fagn},
we show that the AGN fraction increases with Ly$\alpha$ luminosity
(similar trends have been noted by \citealt{cowie10} and \citealt{nilsson11}).
Breaking each sample into 5 luminosity bins from log $L$$_{{\rm {Ly}}\alpha}$
= 42.5 to 43.4 (used below in the Ly$\alpha$ LF computation) and
requiring EW$_{{\rm r}}$(Ly$\alpha$) $\geq$ 20\AA, we find that
the AGN fraction increases from 6\% to 71\% for the full \textit{GALEX}
data cube sample and from 13\% to 83\% for the deep X-ray subsample.
Given this small but systematic difference, we derive the Ly$\alpha$
LF for our full LAE galaxy sample and for the subsample limited to
regions with deep X-ray data to determine the effect, if any, on our
results. 

In common with other low-redshift samples, our AGN fraction is high
compared to the AGN fractions quoted for higher redshift samples.
(Note that many of the low-redshift samples quote AGN fractions that
exclude sources known to be AGNs based on the UV spectra; our AGN
fraction includes these sources.) However, it is essential to emphasize
that the AGN fraction is strongly dependent on the luminosity range.\textbf{
 }Our LAEs have Ly$\alpha$ luminosities from 10$^{42.2}$ to 10$^{43.6}$
erg s$^{-1}$. As shown in \citet{nilsson11}, the AGN fraction rapidly
increases over this range and approaches 100\% at Ly$\alpha$ luminosities
above 10$^{43.6}$erg s$^{-1}$.

The high-luminosity LAEs are primarily AGNs, and it is these objects
that are included in the \textit{GALEX} pipeline extractions. Only
with the increased sensitivity of the data cube search do we probe
faint enough to develop large samples of star-forming LAEs. In Figure
\ref{pipe}(a), we show the LAE redshift versus the NUV magnitudes
for our LAE data cube sample (blue diamonds) and for the pipeline
LAE sample found by \citet[][black squares]{cowie10,cowie11} constrained
to the data cubes' 50$'$$\times$50$'$ FOVs and redshift range ($z=0.67-1.16$).
We note that there are seven pipeline LAE galaxies listed within the
CDFS, GROTH, NGPDWS, and COSMOS fields \citep[see Table 2 in][]{cowie11}.
We excluded one pipeline LAE galaxy due to its low redshift $z=0.65$,
and we excluded two pipeline LAE galaxies because they do not fall
within our data cubes' 50$'$$\times$50$'$ FOVs. Although not excluded
for the purposes of this discussion, if we also require EW$_{{\rm {r}}}$(Ly$\alpha$)
$\geq$ 20\AA, then there are only two LAE galaxies that remain in
the pipeline sample. By construction, the \textit{GALEX} pipeline
extractions miss all LAEs fainter than the pipeline magnitude limit
of NUV$\sim22$ (dashed line in Figure \ref{pipe}(a)). Objects classified
as AGNs in any way have their symbols outlined in red. In Figure \ref{pipe}(b),
we show the LAE redshift versus the Ly$\alpha$ luminosity. Directly
below Figures \ref{pipe}(a) and \ref{pipe}(b), we show these Figures
again but with all AGNs removed. We find that the AGN fraction reaches
100\% at a Ly$\alpha$ luminosity of $\sim10^{43.5}$ erg s$^{-1}$.
This is also the Ly$\alpha$ luminosity where the pipeline extraction
begins to miss sources. While there is some luminosity overlap between
the data cube and pipeline samples, we find that our data cube search
is necessary to obtain a large LAE galaxy sample.

\section{Luminosity function}

We computed the Ly$\alpha$ LF of the combined CDFS, GROTH, NGPDWS,
and COSMOS LAE galaxy samples in the redshift range $z=0.67-1.16$
using the $1/V$ technique \citep{felten76}. We only included sources
that were not classified as AGN in any way and that have EW$_{{\rm {r}}}$(Ly$\alpha$)$\geq20$
\AA. With these criteria, the sample includes two LAE galaxies, which
appeared in the pipeline sample of \citet[][see their Table 2]{cowie11}.
The rest (58) come from our data cube sample and are not found in
the \textit{GALEX} pipeline extractions. 

\begin{deluxetable}{ccc} 
\tablecolumns{3} 
\tablewidth{0pc} 

\tablecaption{Luminosity function minimum Ly$\alpha$ flux and corresponding completeness limit} 
\tablehead{ \colhead{\textit{GALEX} Field} & \colhead{$f_{\rm{Ly}\alpha}$ (erg cm$^{-2}$ s$^{-1}$)} & \colhead{Completeness}\\ 
\colhead{(1)} & \colhead{(2)} & \colhead{(3)}} 
\startdata 
CDFS & 8.5$\times$10$^{-16}$ & $26.4\%$\\ 
GROTH & 5.2$\times$10$^{-16}$ & $1.6\%$\\ 
NGPDWS & 10.9$\times$10$^{-16}$ & $24.6\%$\\ 
COSMOS & 24.1$\times$10$^{-16}$ & $62.3\%$\\ 
\enddata 
\label{flimit}
\end{deluxetable}

To compute the LF, we divided our Ly$\alpha$ survey into 20 samples
(5 samples per field), where the sample $j$ covers a Ly$\alpha$
flux range $f_{1j}\leq f\leq f_{2j}$ and a solid angle area $\omega_{j}$
(the masking of bright sources alters $\omega_{j}$ from one \textit{\small GALEX}
50$'$$\times$50$'$ field to another). For each field, we took the
difference between the minimum and maximum observed Ly$\alpha$ flux
and divided this into 5 flux bins of equal size (see Table \ref{flimit}
for minimum Ly$\alpha$ flux values and corresponding completeness
limits). The sampling rate $S_{j}$ (the fraction of LAE galaxies
in the given flux range that were observed) was estimated with our
completeness simulations. The effective area of a sample is $\Omega_{j}=\omega_{j}S_{j}$.
The accessible volume of the $i^{th}$ LAE galaxy in sample $j$ is

\[
V_{ij}=\Omega_{j}\left(V\left(z_{max}^{ij}\right)-V\left(z_{min}^{ij}\right)\right),
\]

\noindent where the comoving volume is calculated at the highest and
lowest redshifts at which the $i^{th}$ LAE galaxy remains both within
sample $j$'s Ly$\alpha$ flux range $f_{1j}\leq f\leq f_{2j}$ and
within the redshift range $z=0.67-1.16$. The total accessible volume
of the $i^{th}$ LAE galaxy is 

\[
V_{i}=\sum_{j=1}^{20}V_{ij}\:,
\]

\noindent and the Ly$\alpha$ LF in the luminosity range $L_{1}\leq L_{i}\leq L_{2}$
and redshift range $z=0.67-1.16$ is

\[
LF=({\rm {log}(}L_{2})-{\rm {log}}(L_{1}))^{-1}\sum_{\{i:L_{1}\leq L_{i}\leq L_{2}\}}1/V_{i}\:.
\]

In Figure \ref{lf}(a), we show our raw LF (black open circles) along
with our LF corrected for incompleteness (black solid circles). Error
bars are $\pm$1$\sigma$ Poisson errors. In Table \ref{tlf} Columns
1, 2, and 3, we list the luminosity bins, the number of LAE galaxies
per bin, and the computed luminosity function values corrected for
incompleteness. We have tested that these results are not dependent
on our method of constructing flux bins. Alternatively, for each field,
we generated 5 flux bins of equal size within the range of $f_{thres}$
to the maximum observed Ly$\alpha$ flux, where $f_{thres}$ is the
Ly$\alpha$ flux corresponding to 50\% completeness (see Figure \ref{compl}).
This method reduces the dependence on our completeness simulations
but does not significantly change our results. \begin{deluxetable}{ccc} 
\tablecolumns{3} 
\tablewidth{0pc} 

\tablecaption{Cumulative Luminosity Function} 
\tablehead{ 
\colhead{log$(L)$ bin} & \colhead{N} & \colhead{$\frac{1}{\Delta \rm{log}(L)}\sum1/V$}\\ 
\colhead{(erg s$^{-1}$)} & \colhead{} & \colhead{(Mpc$^{-3}$ log(L)$^{-1}$)}\\ 
\colhead{(1)} & \colhead{(2)} & \colhead{(3)}} 
\startdata
42.468 - 42.655 &           17 & 66.3$\pm$16.8$\times 10^{-6}$\\ 
42.655 - 42.842 &           13 & 23.6$\pm$6.7$\times 10^{-6}$\\ 
42.842 - 43.029 &           15 & 14.9$\pm$3.9$\times 10^{-6}$\\ 
43.029 - 43.216 &            8 & 5.7$\pm$2.0$\times 10^{-6}$\\ 
43.216 - 43.403 &            7 & 4.3$\pm$1.6$\times 10^{-6}$\\ 
\enddata 
\label{tlf}
\end{deluxetable} 

We may correct for any remaining AGNs in our sample by restricting
our field to regions with deep X-ray data. This removes the NGPDWS
field and restricts the area of the remaining 50$'\times$50$'$ \textit{GALEX}
fields but ensures a robust AGN classification. In Figure \ref{lf}(b),
we show our X-ray data limited LF (black circles). Comparing this
LF to the LF computed from the full LAE galaxy sample, we find that
all points are consistent within 1$\sigma$ error bars.

We find that our LAE galaxy LF is consistent with the $z\sim1$ LAE
galaxy LF computed in BCW12. BCW12 based their LF on $\sim20$ LAE
galaxies found in one of the four \textsl{GALEX} fields contained
in this study. We have increased the sample size by a factor of three.
Furthermore, we have searched for AGNs in the BCW12 LAE galaxy sample
(none found) and removed spurious sources (3 BCW12 LAE candidates)
with our newly obtained optical spectroscopic data.

\section{Distribution of Equivalent Widths}

\begin{figure}

\includegraphics[bb=130bp 80bp 720bp 495bp,clip,angle=180,width=8.5cm]{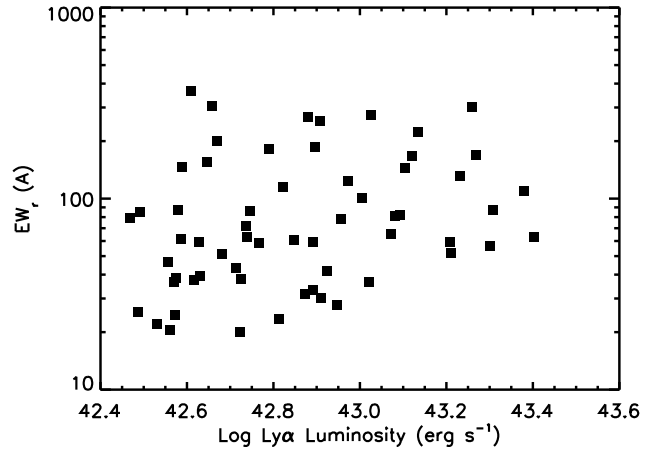}\caption{Rest-frame EWs vs.\ $\log$ Ly$\alpha$ luminosity for the full sample.}
\label{ew_lalum}

\end{figure}

\begin{figure}
\includegraphics[bb=140bp 80bp 700bp 500bp,clip,angle=180,width=8.5cm]{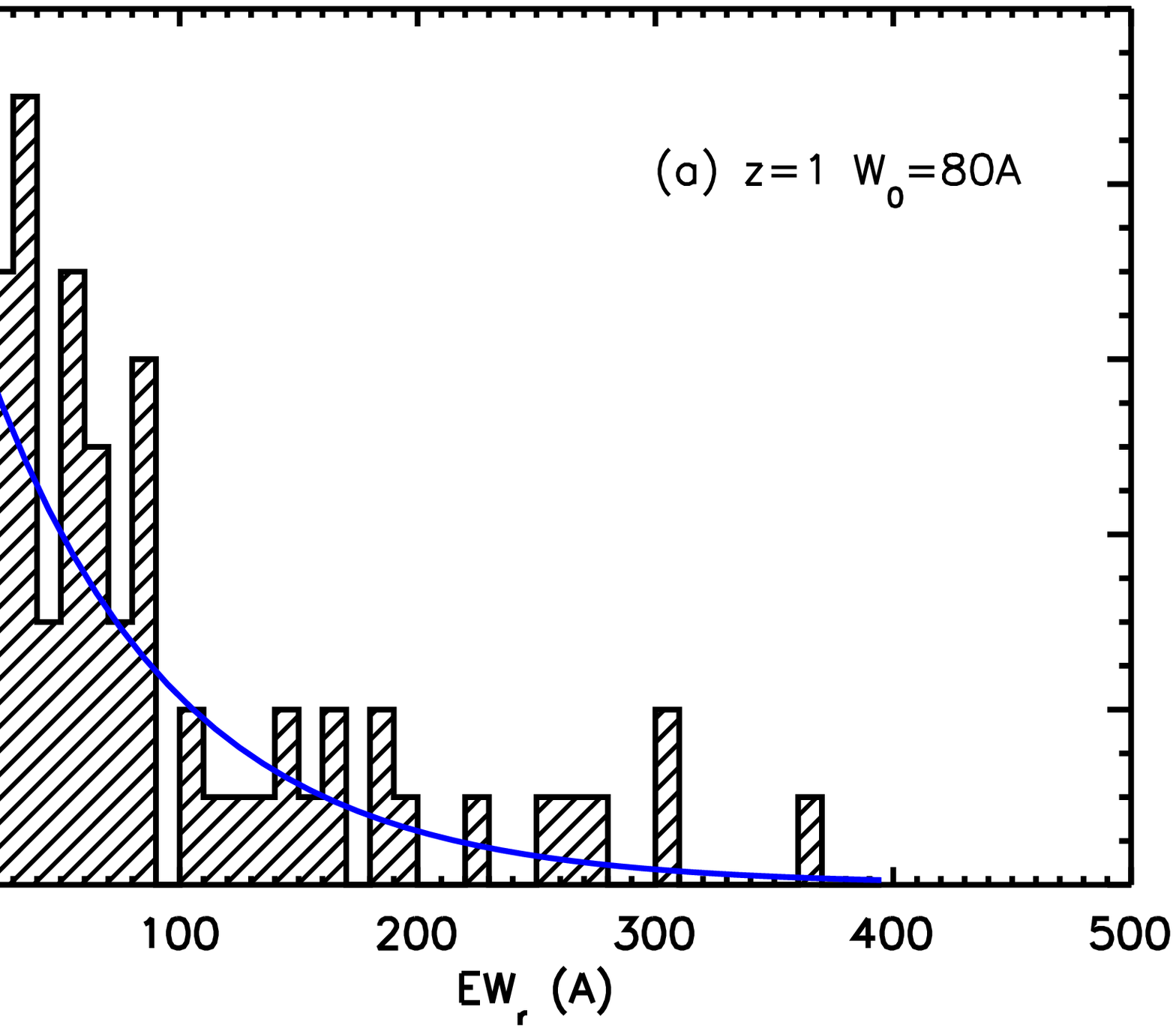}

\includegraphics[bb=140bp 80bp 700bp 500bp,clip,angle=180,width=8.5cm]{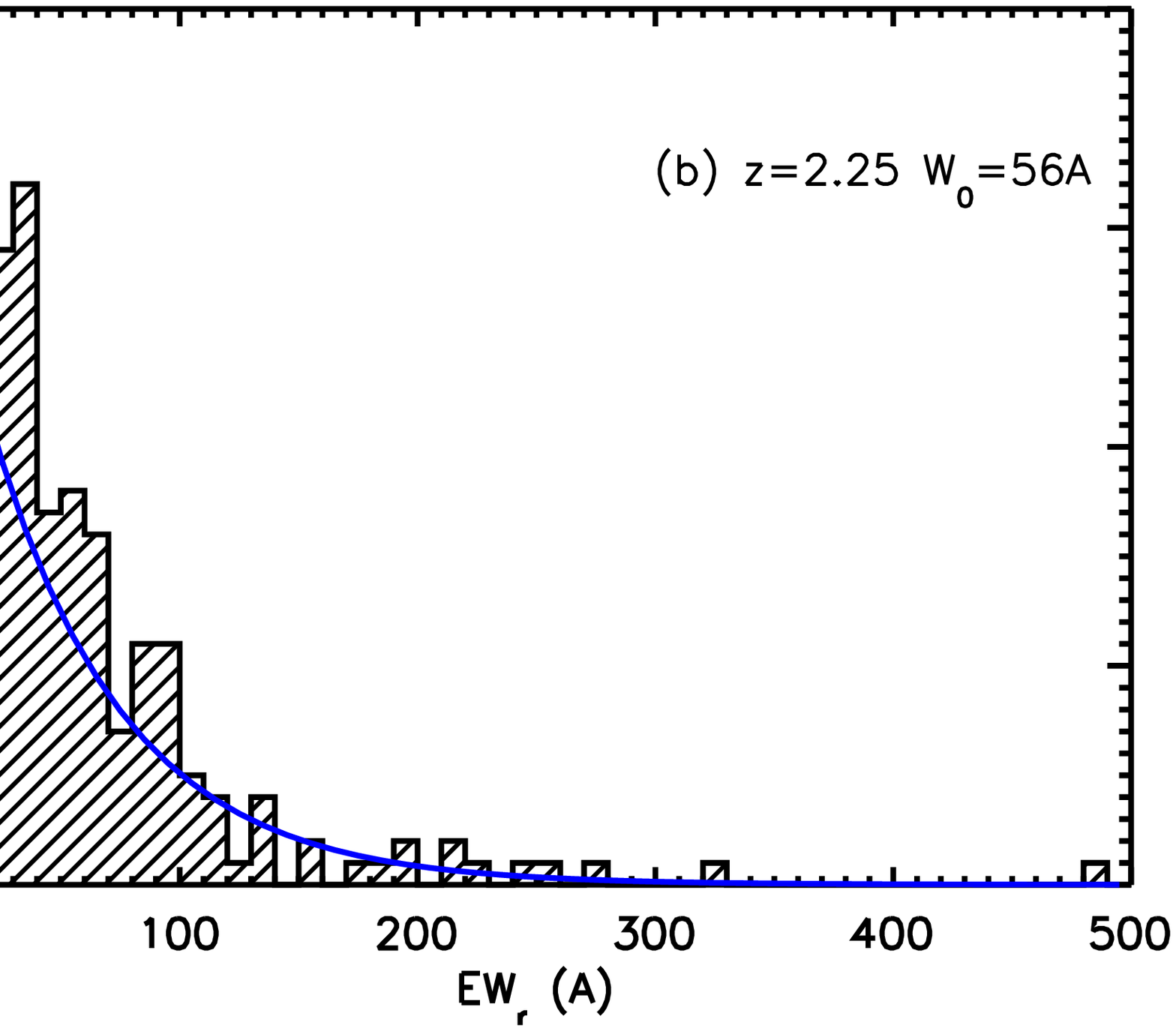}

\includegraphics[bb=140bp 80bp 700bp 500bp,clip,angle=180,width=8.5cm]{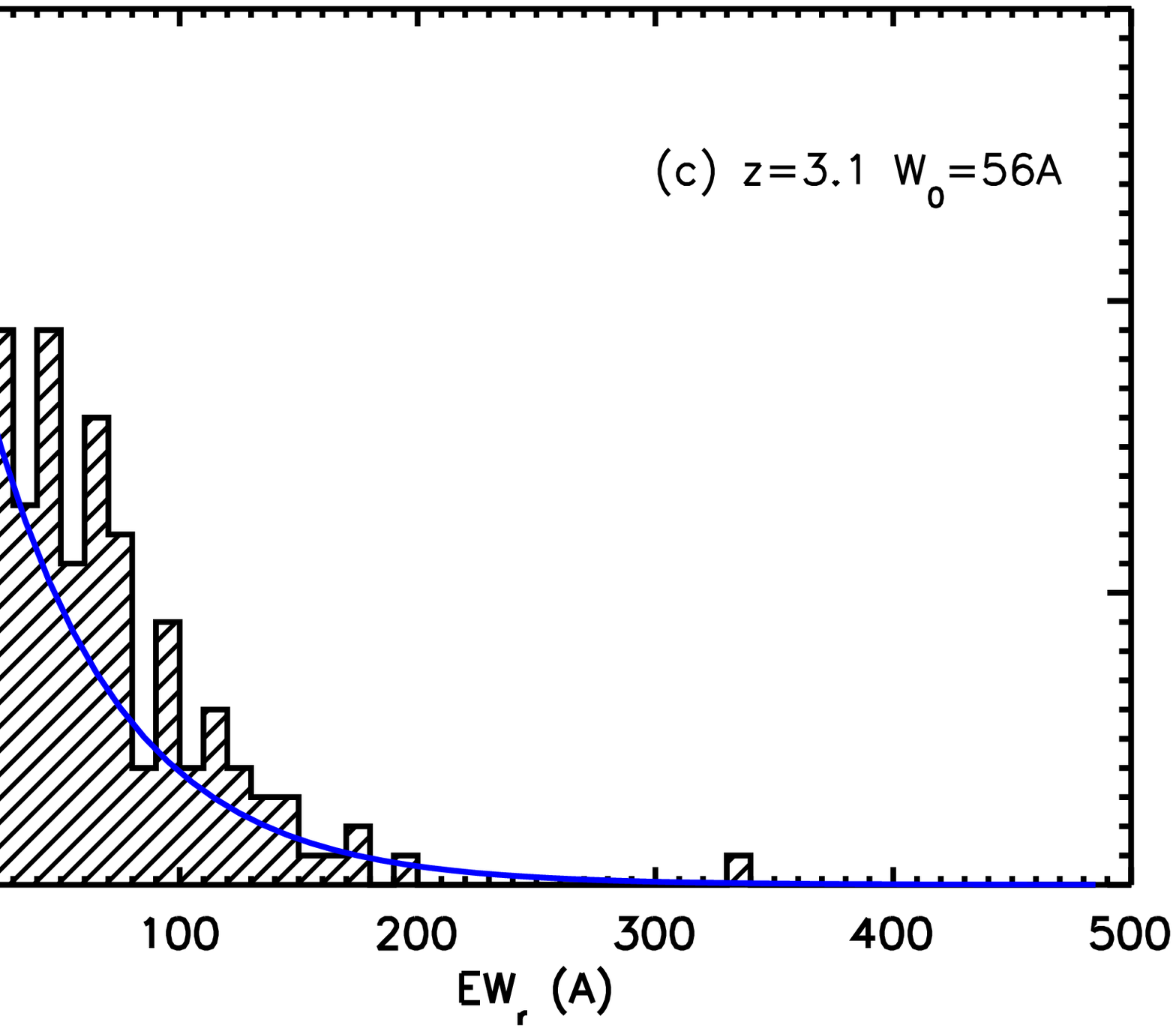}

\caption{Histograms of the rest-frame EWs in 10~\AA\ bins. \textbf{(a)} The
present sample at $z\sim1$. \textbf{(b)} The \citet{nilsson09} sample
at $z=2.25$. \textbf{(c)} The \citet{ciardullo12} sample at $z=3.1$.
In each case, the blue curve shows the exponential distribution computed
using a maximum likelihood procedure. The rest-frame scale length
is 80$\pm$10~\AA\ for $z\sim1$, 56$\pm$7~\AA\ for $z=2.25$,
and 56$\pm$7~\AA\ for $z=3.1$.}

\label{ew_dists}

{\footnotesize (A color version of this figure is available in the online journal.)}
\end{figure}

In Figure~\ref{ew_lalum}, we show our rest-frame EWs versus redshift.
Because we have a flux-limited sample, the distribution of these EWs
should be directly comparable to narrowband Ly$\alpha$ selected samples
at higher redshifts. In Figure~\ref{ew_dists}(a), we show the distribution
of these EWs for the full sample.

\begin{figure}
\includegraphics[bb=130bp 80bp 720bp 495bp,clip,angle=180,width=8.5cm]{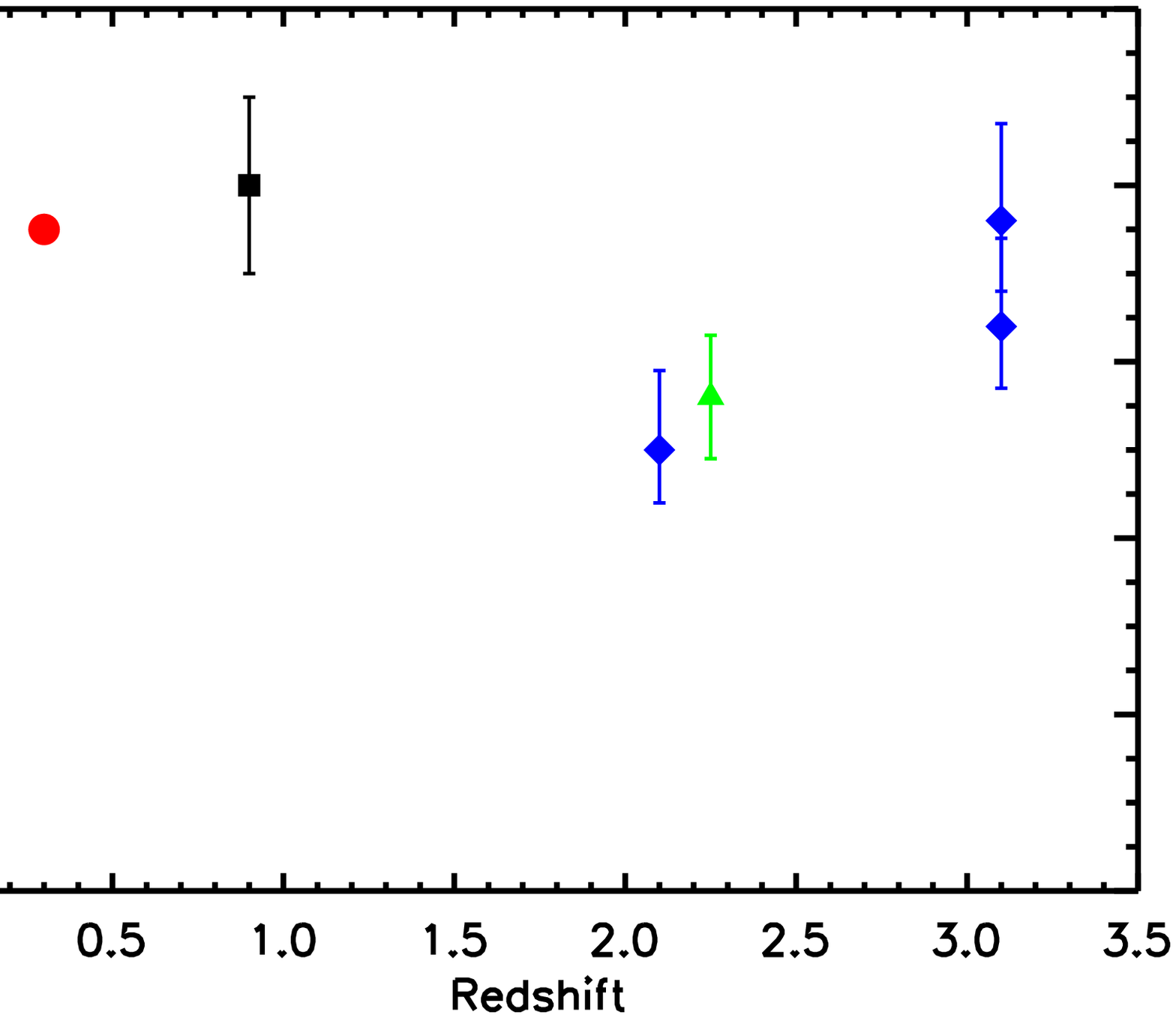}\caption{Maximum likelihood estimates of the rest-frame EW scale length from
the present $z\sim1$ sample (black square), the $z=2.25$ \citet{nilsson09}
sample (green triangle), and the $z=2.1$ \citet{guaita10}, $z=3.1$
\citet{gronwall07}, and $z=3.1$ \citet{ciardullo12} samples (blue
diamonds; values quoted from Ciardullo et al.) vs.\ redshift. The
red circle shows the scale length from the $z=0.3$ {\em GALEX\/}
LAE sample of \citet{cowie10} after correction for their continuum
selection, as described in their Section~5.4.}

\label{ew_z}

{\footnotesize (A color version of this figure is available in the online journal.)}
\end{figure}

The rest-frame EW distribution is normally fit with an exponential
function. For the LAE samples, which are truncated below $20$~\AA,
the maximum likelihood estimate of the scale length is the mean of
the rest-frame EW values minus $20$~\AA. We find a scale length
of $80\pm10$~\AA\ for the full sample, where we computed the error
using the parameterized bootstrap method. Separating the sample by
luminosity gives a scale length of $91\pm14$~\AA\ for sources with
logarithmic Ly$\alpha$ luminosities above 42.8 and $68\pm10$~\AA\
for those with lower luminosities. This is consistent with the distribution
being invariant as a function of Ly$\alpha$ luminosity, as has been
found in the higher redshift samples \citep{ciardullo12}.

We compare our sample at $z\sim1$ with a sample at $z=2.25$ from
\citet{nilsson09} (Figure~\ref{ew_dists}(b)) and a sample at $z=3.1$
from \citet{ciardullo12} (Figure~\ref{ew_dists}(c)). The Nilsson
et al.\ sample has a logarithmic luminosity limit of 42.4~erg~s$^{-1}$,
which is very similar to that of the present sample, while the Ciardullo
et al.\ sample covers 42~erg~s$^{-1}$ to just above 43~erg~s$^{-1}$,
making it slightly fainter, on average. The Nilsson et al.\ and Ciardullo
et al.\ maximum likelihood estimates of the rest-frame EW scale length
are both $56\pm7$~\AA. Nilsson et al.\ also give a least squares
fit value of $56\pm7$~\AA, which is consistent within the errors. 

However, when \citet{ciardullo12} allowed for biases in their sample,
such as those introduced by the filter shape, then their rest-frame
EW scale length rose to $64_{-7}^{+10}$~\AA. When Ciardullo et al.\
analyzed an alternate sample at $z=3.1$ from \citet{gronwall07}
in the same way, they found $76_{-8}^{+11}$~\AA. From the combined
samples, they found $70_{-5}^{+7}$~\AA. Ciardullo et al.\ also
reanalyzed the $z=2.1$ sample from \citet{guaita10} in a self-consistent
way and found $50_{-6}^{+9}$~\AA. The Ciardullo et al.\ $z=2.1$
value is consistent with the \citet{nilsson09} $z=2.25$ value. Based
on an Anderson-Darling test, they concluded that there is a significant
difference between the $z=3.1$ and $z=2.1$ samples. 

The \citet{ciardullo12} analysis emphasizes that differences inherent
in the methodology and fitting procedure are important. In particular,
comparisons between the present sample, where the line fluxes are
computed from the spectra and the continua from the broadband magnitudes,
and samples such as the Ciardullo et al.\ sample, which are based
purely on imaging data, can be tricky \citep[e.g., ][]{zheng13}.
Nevertheless, the present sample appears to indicate that, if there
is indeed a narrowing of the rest-frame EW distribution from $z=3$
to $z=2$, then it has reversed by $z=1$.

In Figure~\ref{ew_z}, we show the rest-frame EW scale lengths measured
from the maximum likelihood procedures for the present sample (black
square), the $z=2.25$ \citet{nilsson09} sample (green triangle),
and the $z=2.1$ \citet{guaita10}, $z=3.1$ \citet{ciardullo12},
and $z=3.1$ \citet{gronwall07} samples (blue diamonds; values quoted
from Ciardullo et al.) While there may be a drop at $z\sim2$, overall
the scale length shows little variation with redshift. This is in
sharp contrast to the analysis of \citet[][]{zheng13}, who found
a redshift evolution of $(1+z)^{1.7}$. The Zheng et al.\ result
appears to be due primarily to their misinterpretation of the scale
length given by \citet{cowie10} for their $z=0.3$ {\em GALEX\/}
LAE sample. Zheng et al.\ used $23.7\pm2.2$, which was what Cowie
et al.\ gave for their continuum selected sample. However, as Cowie
et al.\ emphasized in their Section~5.4, this needs to be corrected
for missing high EW sources before it can be compared with Ly$\alpha$
luminosity selected samples. In Figure~\ref{ew_z}, we show the Cowie
et al.\ corrected scale length of 75~\AA\ (red circle). (Because
of the large systematic errors, we do not show an error bar.) This
is fully consistent with there being a constant value of the scale
length with redshift. 

\begin{figure}
\includegraphics[bb=90bp 50bp 670bp 520bp,clip,angle=180,scale=0.4]{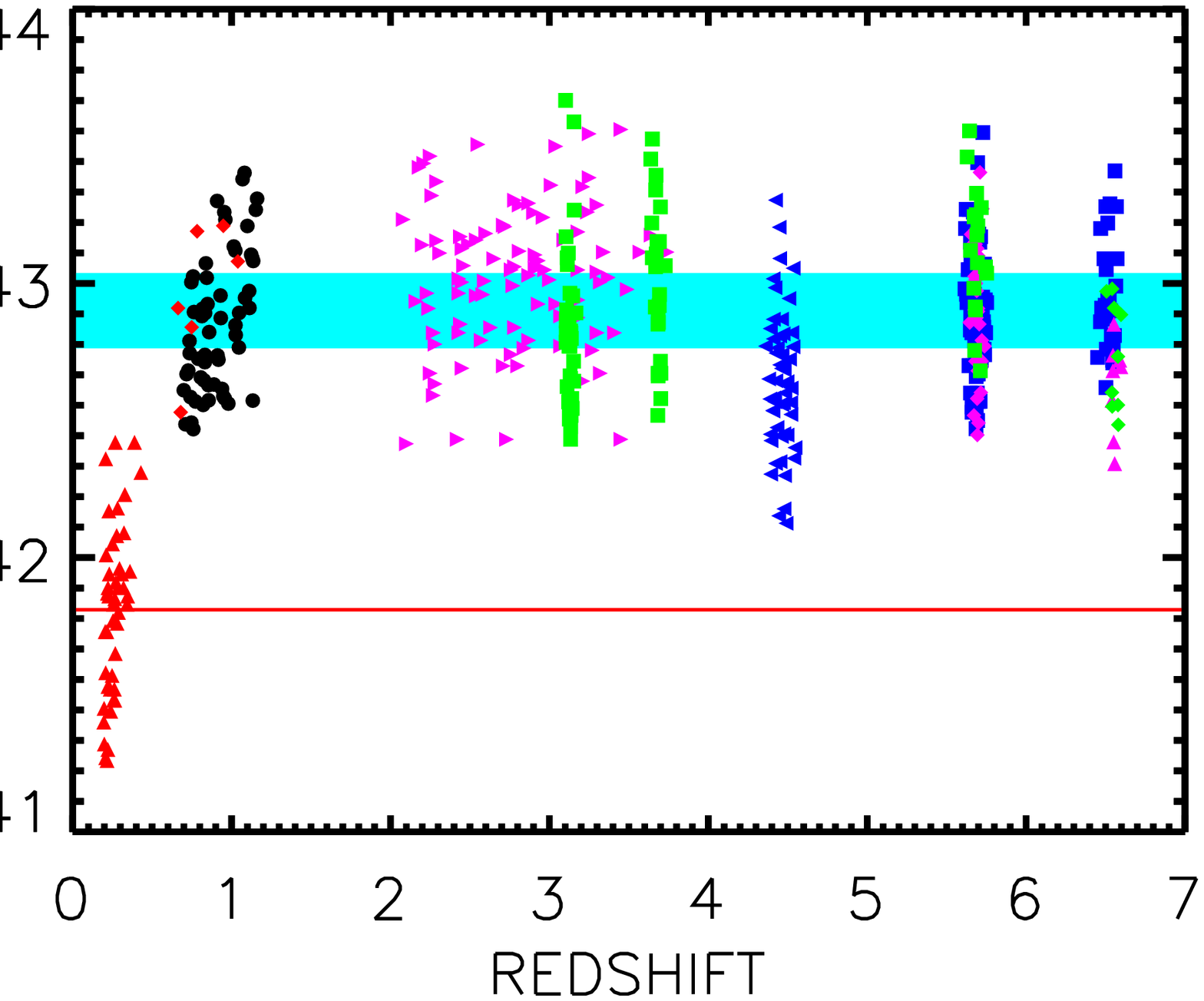}\caption{Observed Ly$\alpha$ luminosity vs. redshift for LAE galaxies with
EW$_{{\rm {r}}}$(Ly$\alpha$) $\geq$ 20 \AA. We show pipeline \textit{GALEX}
LAEs from \citet[][red triangles and red diamonds]{cowie11}, data
cube \textit{GALEX} LAEs from this study (black circles), $z=1.9-3.8$
HETDEX LAEs from \citet[][magenta right pointing triangles]{blanc11},
$z=(3.1,\:3.7,\:5.7)$ LAEs from \citet[][green squares]{ouchi08},
$z=4.5$ LAEs from \citet[][blue left pointing triangles]{dawson07},
$z=(5.7,\:6.5)$ LAEs from \citet[][blue squares]{hu10}, $z=5.7$
LAEs from \citet[][magenta diamonds]{shimasaku06}, and $z=6.5$ LAEs
from \citet[][green diamonds]{taniguchi05} and \citet[][magenta triangles]{kashikawa06}.
The red line shows the $L_{\star}$ derived for the $z\sim0.3$ LF
assuming a slope of $\alpha=-1.6$. The cyan area shows the range
of $L_{\star}$ values derived for the $z=2.85-5.7$ LFs.}

\label{lstar}

{\footnotesize (A color version of this figure is available in the online journal.)}
\end{figure}
\begin{figure}
\includegraphics[bb=75bp 75bp 700bp 520bp,clip,angle=180,scale=0.4]{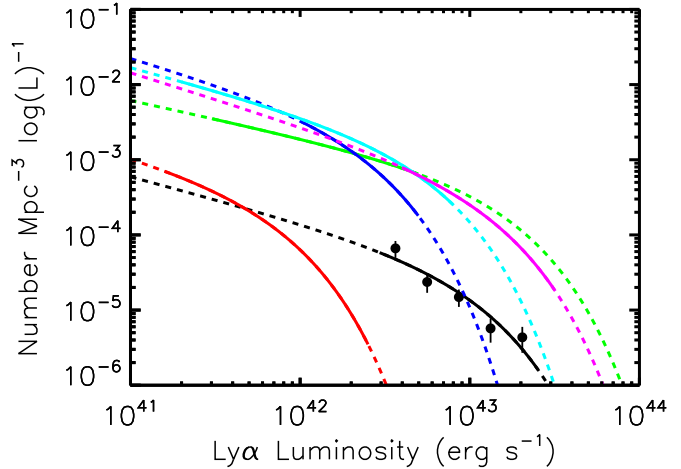}\caption{Derived Ly$\alpha$ luminosity function at $z=0.67-1.16$ for the
LAE galaxies in deep \textit{GALEX} grism fields (black circles from
Figure \ref{lf}(a)) compared to other Ly$\alpha$ luminosity functions
in close redshift proximity. The fitted Schechter functions are shown
as solid curves over the the extent of their observed data points.
The black curve indicates our best fit Schechter function assuming
a fixed slope of $\alpha=-1.6$. The red curve indicates the $z=0.194-0.44$
\citet{cowie10} LF. The blue curve indicates the $z=2.1$ \citet{ciardullo12}
LF. The green curve indicates the $z=2.2$ \citet{hayes10} LF. The
cyan curve indicates the $z=1.95-3$ \citet{cassata11} LF. The magenta
curve indicates the $z=1.9-3.8$ \citet{blanc11} LF.}

\label{complf}

{\footnotesize (A color version of this figure is available in the online journal.)}
\end{figure}
\begin{figure}
\includegraphics[bb=75bp 75bp 700bp 520bp,clip,angle=180,scale=0.4]{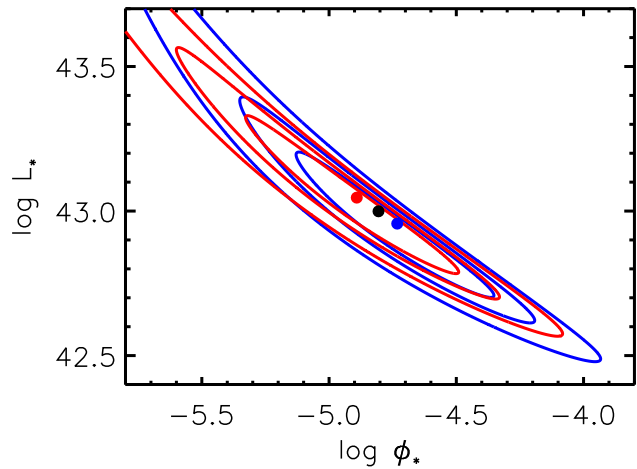}\caption{1$\sigma$, 2$\sigma$, and 3$\sigma$\textbf{\textit{ }}contours
for the $z\sim1$ luminosity function parameters $L_{\star}$ and
$\phi_{\star}$ assuming a faint end slope of $\alpha=-1.5$ (blue)
and $\alpha=-1.7$ (red). The blue, black, and red circles show our
best fit Schechter function results for the full LAE galaxy sample
assuming $\alpha=-1.5$ (blue circle), $\alpha=-1.6$ (black circle),
and $\alpha=-1.7$ (red circle). Regardless of the assumed faint end
slope, log $L_{\star}\phi_{\star}\sim38.2$.}

\label{contour}

{\footnotesize (A color version of this figure is available in the online journal.)}
\end{figure}
\begin{figure}
\includegraphics[bb=75bp 75bp 700bp 520bp,clip,angle=180,scale=0.4]{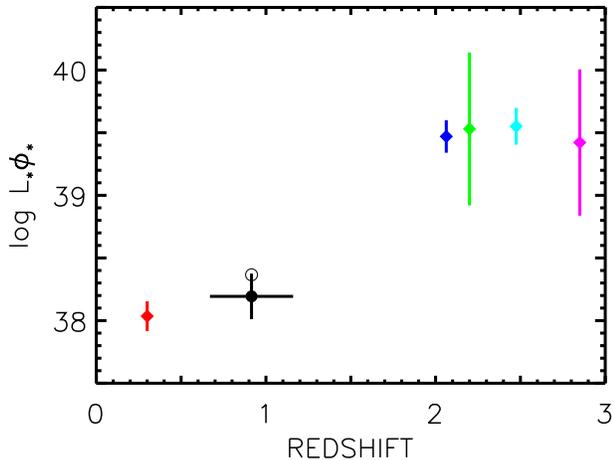}\caption{The product $L_{\star}\phi_{\star}$ vs.\ redshift for the LAE galaxy
LFs shown in Figure \ref{lf}. The black solid circle shows the product
$L_{\star}\phi_{\star}=38.2\pm0.2$ from our best fit $z\sim1$ LAE
galaxy LF. The black open circle shows the product $L_{\star}\phi_{\star}=38.4\pm0.2$
from our fit to the $z\sim1$ LAE galaxy LF limited to regions with
deep X-ray data. Both $z\sim1$ fits assume a faint end slope $\alpha=-1.6$.
Error bars are estimated from the published uncertainties in $L_{\star}$
and $\phi_{\star}$.}

\label{lphi}

{\footnotesize (A color version of this figure is available in the online journal.)}
\end{figure}
\begin{figure}
\begin{centering}
\includegraphics[bb=75bp 75bp 700bp 520bp,clip,angle=180,scale=0.4]{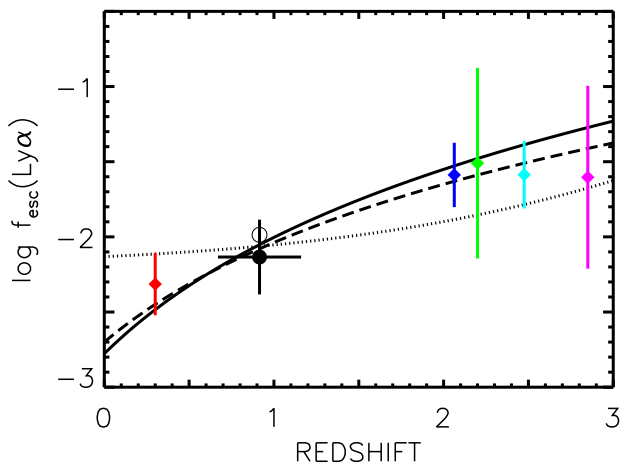}\caption{Sample averaged Ly$\alpha$ escape fractions for all LFs presented
in Figure \ref{lf}. The black solid circle shows $f_{esc}({\rm {Ly}}\alpha)=0.7\%\pm0.4\%$
computed from our best fit $z\sim1$ LF. The black open circle shows
$f_{esc}({\rm {Ly}}\alpha)=1.0\%\pm0.6\%$ computed from our fitted
LF limited to regions with deep X-ray data. Diamonds (from low redshift
to high) indicate $f_{esc}({\rm {Ly}}\alpha)$ derived from \citet{cowie10},
\citet{ciardullo12}, \citet{hayes10},  \citet{cassata11}, and \citet{blanc11}.
Error bars are estimated from the published uncertainties in $L_{\star}$
and $\phi_{\star}$. The solid and dashed black curves are from \citet{hayes11}
and \citet{blanc11}, respectively, and show the best fit power law
to $f_{esc}({\rm {Ly}}\alpha)$ data. The dotted black curve shows
the best fit transition model from \citet{blanc11}. The calculated
points and the selected curves have not been corrected for IGM absorption,
which should be negligible at $z\sim1$ and within the 1$\sigma$
uncertainties at $z<3$.}

\par\end{centering}

\begin{centering}
\label{fesc}
\par\end{centering}

\centering{}{\footnotesize (A color version of this figure is available in the online journal.)}
\end{figure}

\section{Discussion}

In Figure \ref{lstar}, we show the dramatic increase in the observed
Ly$\alpha$ luminosities to a redshift of $z\sim1$ followed by relatively
no luminosity evolution from $z\sim1$ to $z\sim6$. This large
luminosity boost was previously noted by \citet{cowie11} with a sample
of $\sim5$ LAEs at a redshift of $z\sim1$ (see Figure \ref{lstar},
red diamonds). We have now increased the sample size of $z\sim1$
LAEs to 60 galaxies (see Figure \ref{lstar}, black circles). Our
relatively nearby $z\sim1$ sample will facilitate the study of LAEs
with luminosities analogous to high-redshift LAEs. In Paper II, we
will use the available spectroscopic and imaging data to constrain
the properties of our sample and to serve as a baseline for studies
of higher redshift LAEs. 

In order to quantify the evolution in $L_{\star}$, we have held
the slope $\alpha$ fixed and found the best \citet{schechter76}
function fit to our data. We assume a slope of $\alpha=-1.6$, which
allows us to compare directly our results to the LAE galaxy LF at
$z\sim0.3$ \citep[$\alpha=-1.6$, log $L_{\star}=41.81\pm0.09$, log $\phi_{\star}=-3.77\pm0.08$;][]{cowie10}.
We obtain log $L_{\star}=43.0\pm0.2$ and log $\phi_{\star}=-4.8\pm0.3$
(see Figure \ref{lf}(a)). Limiting our sample to regions with deep
X-ray data, we obtain log $L_{\star}=42.8\pm0.2$ and log $\phi_{\star}=-4.5\pm0.3$
(see Figure \ref{lf}(b)). 

Both our $L_{\star}(z\sim1)$ values are roughly consistent with $L_{\star}$
values found for LAE galaxy LFs at $z\sim3$ (log $L_{\star}=42.7$,
\citealt{gronwall07}; log $L_{\star}=42.8$, \citealt{ouchi08};
log $L_{\star}=42.8$, \citealt{ciardullo12}) and imply an $L_{\star}$
increase by a factor of $\sim15$ from $z\sim0.3$ to $z\sim1$ (from
log $L_{\star}(z\sim0.3)=41.81\pm0.09$ to log $L_{\star}(z\sim1)=43.0\pm0.2$).
A large boost in $L_{\star}$ is in agreement with the evolution seen
in other LFs tracing star-forming galaxies. Both H$\alpha$ and UV
LFs are found to be dominated by luminosity evolution over the redshift
range $z\sim0.3-2$. \citet{sobral13} investigated H$\alpha$ LFs
and found that $L_{\star}({\rm H}\alpha)$ increases by a factor of
10 from $z\sim0$ to $z\sim2$. \citet{oesch10} investigated UV LFs
and found that $L_{\star}({\rm UV})$ increases by a factor of $\sim16$
from $z\sim0$ to $z\sim3$.

However, our fitted Schechter parameters indicate a decrease in $\phi_{\star}$
from $z\sim0.3$ to $z\sim1$ (from log $\phi_{\star}(z\sim0.3)=-3.77\pm0.08$
to log $\phi_{\star}(z\sim1)=-4.8\pm0.3$).  It is only beyond $z\sim1$
that there a significant increase in $\phi_{\star}$. Referring again
to other LFs tracing star-forming galaxies, we note that, unlike $L_{\star}$,
$\phi_{\star}$ does not increase monotonically over the redshift
range $z\sim0.3-2$. Within the uncertainties of their measurements,
both \citet[][for $\phi_{\star}({\rm H}\alpha)$]{sobral13} and \citet[][for $\phi_{\star}({\rm UV})$]{oesch10}
found very little evidence for $\phi_{\star}$ evolution from $z\sim0$
to $z\sim2$. As discussed in more detail below, we find that our
best fit Schechter functions with higher $L_{\star}$ values tend
to have lower $\phi_{\star}$ values. Given this issue and the poorly
constrained faint end slope, a clearer picture of the evolution of
individual Schechter function parameters awaits a $z\sim1$ LAE survey
probing lower Ly$\alpha$ luminosities. Such a survey is not possible
with currently available telescopes but would allow stronger constraints
on the $L_{\star}(z\sim1)$ and $\phi_{\star}(z\sim1)$ values.

In Figure \ref{complf}, we compare our $z\sim1$ LF computed from
the full LAE galaxy sample to other Ly$\alpha$ luminosity functions
in close redshift proximity. We show the fitted Schechter functions
as solid lines over the the extent of their observed data points.
For the $z\sim2$ Schechter function parameters, the log $L$$_{\star}$
values range from 42.3 to 43.2, the log $\phi_{\star}$ values range
from $-3.7$ to $-2.9$, and the $\alpha$ values range from $-1.7$
to $-1.5$ (see the blue, green, cyan, and magenta lines in Figure
\ref{complf}). Due to the differing survey volumes and fitting assumptions,
it is still possible that all $z\sim2$ data are in overall agreement
\citep{blanc11,ciardullo12}. 

The Schechter function parameters $\phi_{\star}$ and $L_{\star}$
are known to be strongly correlated, while assuming different values
of $\alpha$ tends to leave the product $L_{\star}\phi_{\star}$ unchanged.
We show this for our Schechter function fit in Figure \ref{contour}.
For these reasons, we consider the product $L_{\star}\phi_{\star}$
and the luminosity density, which is proportional to $L_{\star}\phi_{\star}$,
to be more reliable than the values of individual Schechter parameters.
In Figure \ref{lphi}, we show that a coherent picture emerges by
comparing our results to the $L_{\star}\phi_{\star}$ values found
in the literature. As noted by \citet[][]{blanc11}, there is a trend
(in agreement with our uncertainty contours) for the $z\sim2$ Schechter
function fits with higher $L_{\star}$ values to have lower $\phi_{\star}$
values. $ $Thus, the $z\sim2$ Schechter functions --- which have
$L_{\star}$ values that differ by about an order of magnitude ---
are found to have consistent $L_{\star}\phi_{\star}$ values. The
product $L_{\star}\phi_{\star}$ increases by a factor of $\sim1.5$
from $z\sim0.3$ to $z\sim1$ followed by an increase by a factor
of $\sim20$ from $z\sim1$ to $z\sim2$. 

The observed Ly$\alpha$ luminosity density provides an estimate of
the total amount of Ly$\alpha$ light emitted by galaxies per unit
volume:
\[
\rho_{Ly\alpha}^{obs}=\int L\Phi(L)dL,
\]

\noindent where $\Phi(L)$ is a Ly$\alpha$ Schechter function. Integrating
this expression from $0$ to $+\infty$ gives

\[
\rho_{Ly\alpha}^{obs}=L_{\star}\phi_{\star}\Gamma(\alpha+2),
\]

\noindent where $\Gamma$ is the Gamma function. The main source of
uncertainty in this expression comes from the poorly constrained faint
end slope $\alpha$. Taking the extreme values for the $z\sim2$ Schechter
functions, $\alpha=-1.7$ and $\alpha=-1.5$, we find that $\Gamma$
changes by a factor of 1.7. Modulo this factor, the evolution of the
product $L_{\star}\phi_{\star}$ seen in Figure \ref{lphi} from $z\sim0.3$
to $z\sim2$ results from the change in observed Ly$\alpha$ luminosity
density. 

To first order, the intrinsic production of Ly$\alpha$ photons in
a galaxy is proportional to the SFR \citep{kennicutt98,schaerer03}.
Thus, an overall increase in the Ly$\alpha$ luminosity density from
$z\sim0.3$ to $z\sim2$ is expected due to the well established increase
in the cosmic SFR density.  However, observed Ly$\alpha$ luminosities
will significantly deviate from intrinsic Ly$\alpha$ luminosities
due to resonant scattering of Ly$\alpha$ photons by neutral hydrogen.

To quantify this, we calculate the $z\sim1$ volumetric Ly$\alpha$
escape fraction, $f_{esc}({\rm {Ly}}\alpha)$, and compare its value
to other measurements taken at nearby redshifts. The volumetric Ly$\alpha$
escape fraction measures the fraction of Ly$\alpha$ photons that
escape from the survey volume. We emphasize that its value will be
systematically lower than the Ly$\alpha$ escape fractions derived
for individual LAEs. For example, \citet{blanc11} find a $\sim30\%$
median Ly$\alpha$ escape fraction for their $z=1.9-3.8$ LAE sample
but only a $\sim3\%$ volumetric Ly$\alpha$ escape fraction.

The volumetric escape fraction is defined as the observed Ly$\alpha$
luminosity density divided by the intrinsic Ly$\alpha$ luminosity
density,

\[
f_{esc}(Ly\alpha)=\rho_{Ly\alpha}^{obs}/\rho_{Ly\alpha}^{int},
\]
where the intrinsic Ly$\alpha$ luminosity density is found by taking
a measure of the cosmic SFR density and then converting SFR to Ly$\alpha$
luminosity,

\[
\rho_{{\rm {Ly}}\alpha}^{int}=\rho_{SFR}(8.7)(1.26\times10^{41}).
\]

\noindent The conversion factors come from the assumption of case
B recombination \citep{brocklehurst71} and the SFR to H$\alpha$
\citet{kennicutt98} relation. For consistency with \citet{blanc11},
we integrate the Ly$\alpha$ LFs down to a $2.66\times10^{41}$erg
s$^{-1}$ limit and obtain cosmic SFR density measurements from \citet{bouwens10}.
\citet{blanc11} estimate that the choice of the luminosity integration
limit may alter the computed luminosity density by at most $60\%$
\citep[for further discussion of this issue see][]{hayes11}. \citet{bouwens10}
derive cosmic SFR densities from extinction corrected rest-frame UV
LFs. We do not attempt to correct for IGM absorption, which should
be negligible for our $z\sim1$ sample. 

Both \citet{hayes11} and \citet{blanc11} have compiled measurements
from various Ly$\alpha$ LF studies spanning a redshift range of $z=0.3-7.7$
to determine the evolution of $f_{esc}({\rm {Ly}}\alpha)$. Both groups
fit power laws to these data and find evidence for a rapidly increasing
$f_{esc}({\rm {Ly}}\alpha)$, which approaches 100$\%$ at high redshifts
($z\sim9$). 

We find a $z\sim1$ $f_{esc}({\rm {Ly}}\alpha)$ of $0.7\%\pm0.4\%$.
In Figure \ref{fesc}, we show that the $f_{esc}({\rm {Ly}}\alpha)$
computed from our best fit $z\sim1$ Schechter function is consistent
with the $f_{esc}({\rm {Ly}}\alpha)$ value of $\sim0.9\%$ interpolated
from the power law fits of \citet{hayes11} and \citet{blanc11}.
In Figure \ref{fesc}, we also show $f_{esc}({\rm {Ly}}\alpha)=1.0\%\pm0.6\%$
(open circle) computed from our fitted LF limited to regions with
deep X-ray data.  \citet{hayes11} argue that the rising trend
in the volumetric Ly$\alpha$ escape fraction with increasing redshift
is consistent with expectations due to the general decline of dust
content in star-forming galaxies. We have now constrained the $f_{esc}({\rm {Ly}}\alpha)$
from $z\sim0.3$ to $z\sim2$ and find results consistent with this
inferred trend.

\section{Summary}

We presented a catalog of 135 candidate $z\sim1$ LAEs. We obtained
optical spectral data for 90\% of our sample. We found that only $\sim$13\%
of our sample are either spurious, stars, or strong C{\small IV$\lambda$}1549
emitters. Combining these optical data with the UV spectra and X-ray
imaging data, we found that 49\% of our Ly$\alpha$ emitters are AGNs.
Eliminating AGNs and LAEs with  EW$_{{\rm {r}}}$(Ly$\alpha$)$<$20
\AA~ gives a final sample of 60 star-forming LAEs, which we used
to compute the $z\sim1$ Ly$\alpha$ LF and the $z\sim1$ Ly$\alpha$
volumetric escape fraction ($f_{esc}({\rm {Ly}}\alpha)=0.7\%\pm0.4\%$).
We note that no improved $z\sim1$ Ly$\alpha$ LFs are possible for
the foreseeable future due to the lack of UV telescopes. Our best
fit LF implies a significant increase in $L_{\star}$ between $z\sim0.3$
and $z\sim1$ without much change in the luminosity density. This
requires a rapid increase in $\phi_{\star}$ between $z\sim1$ and
$z\sim2$. It is clear that the intrinsic properties of the LAEs must
be changing rapidly in the $z\sim1-2$ interval, resulting in a rapid
increase in the Ly$\alpha$ escape fraction. Our cataloged sources
offer the best opportunity to study emitters with luminosities comparable
to LAEs found in the early universe. With the optical data in hand,
Paper II will compare our sample's physical properties to a UV-selected
control sample.

\acknowledgements{We would like to thank the referee for their critical reading of
the paper and useful suggestions for improving it. We gratefully acknowledge
support from a NASA Keck PI Data Award, administered by the NASA Exoplanet
Science Institute, a Wisconsin Space Grant Consortium Graduate Fellowship,
and a Sigma Xi Grant in Aid of Research Award (I.G.B.W.). We also
appreciate support from the University of Wisconsin Research Committee
with funds granted by the Wisconsin Alumni Research Foundation and
the David and Lucile Packard Foundation (A.J.B.), as well as from
NSF grant AST-0709356 (L.L.C.). We thank Michael Cooper for supplying
spectra from the Arizona CDFS Environment Survey (ACES). This research
has made use of the NASA/IPAC Extragalactic Database (NED), which
is operated by the Jet Propulsion Laboratory, California Institute
of Technology, under contract with the National Aeronautics and Space
Administration.}



\clearpage \begin{landscape} 
\begin{deluxetable}{ccccccccccccccc} 
\tabletypesize{\scriptsize} 
\setlength{\tabcolsep}{0.02in} 
\tablecolumns{15} 
\tablewidth{0pc} 
\tablecaption{Emission-line Sample: CDFS-00} 
\tablehead{ \colhead{Name}    &  \colhead{R.A.} &   \colhead{Decl.} & \colhead{NUV} & \colhead{FUV} & \colhead{$z_{\rm{galex}}$} & \colhead{log $L$(Ly$\alpha$)} & \colhead{EW$_{\rm{r}}$(Ly$\alpha$)} & \colhead{log $f^{a}_{2-8~  \rm{keV}}$} & \colhead{Type(UV)} & \colhead{R.A.(opt)} &  \colhead{Decl.(opt)} & \colhead{Offset} & \colhead{$z_{\rm{opt}}$} & \colhead{Type(opt)}\\ 
\colhead{} & \colhead{(J2000.0)}   & \colhead{(J2000.0)} & \colhead{(AB)} & \colhead{(AB)} & \colhead{} & \colhead{(erg s$^{-1}$)} & \colhead{(\AA)} & \colhead{(erg cm$^{-2}$s$^{-1}$)} & \colhead{} & \colhead{(J2000.0)}   & \colhead{(J2000.0)} & \colhead{(arcsec)} & \colhead{} & \colhead{}\\ \colhead{(1)} & \colhead{(2)} & \colhead{(3)} & \colhead{(4)} & \colhead{(5)} & \colhead{(6)} & \colhead{(7)}  & \colhead{(8)}  & \colhead{(9)} & \colhead{(10)} & \colhead{(11)} & \colhead{(12)} & \colhead{(13)} & \colhead{(14)} & \colhead{(15)}} \startdata GALEX033111-281725\tablenotemark{i} &        52.796374 &       -28.290444 & 23.55 & 27.50 & 1.168 & 43.61 & 281$\pm$17 & \nodata & \nodata & 52.796253 & -28.290321 & 0.6 & 1.164\tablenotemark{b}\phm{d}\phm{a} & AGN\\ GALEX033112-281517\phn &        52.800377 &       -28.254936 & 23.02 & 28.01 & 1.162 & 43.31 & 87$\pm$5 & \nodata & \nodata & 52.800663 & -28.255022 & 1.0 & 1.161\tablenotemark{b,c}\phm{d} & \nodata\\ GALEX033359-275326\phn &        53.499160 &       -27.890792 & 22.04 & 24.13 & 1.143 & 43.53 & 63$\pm$2 & \nodata & AGN & 53.499096 & -27.890846 & 0.3 & 1.133\tablenotemark{b}\phm{d}\phm{a} & AGN\\ GALEX033301-273227\phn &        53.255196 &       -27.540939 & 23.45 & -27.75\phm{-} & 1.135 & 43.08 & 82$\pm$7 & \nodata & \nodata & 53.255268 & -27.540880 & 0.3 & 1.136\tablenotemark{b}\phm{d}\phm{a} & \nodata\\ GALEX033146-272942\phn &        52.942574 &       -27.495223 & 22.73 & 25.44 & 1.137 &   &   & \nodata & \nodata & 52.942543 & -27.495167 & 0.3 & (0.629)\tablenotemark{b}\tablenotemark{,g}\phm{a} & \nodata\\ GALEX033348-280216\phn &        53.453649 &       -28.038008 & 23.40 & 25.53 & 1.128 & 43.09 & 82$\pm$8 & \nodata & \nodata & 53.453671 & -28.038050 & 0.2 & 1.130\tablenotemark{b}\phm{d}\phm{a} & \nodata\\ GALEX033230-280804\phn &        53.125683 &       -28.134514 & 22.34 & 23.83 & 1.124 & 43.27 & 47$\pm$3 & \nodata & AGN & \nodata & \nodata & \nodata & \nodata\phm{d}\phm{a} & \nodata\\ GALEX033120-274917\phn &        52.837504 &       -27.821383 & 22.82 & 32.73 & 1.116 & 42.91 & 30$\pm$5 & \nodata & \nodata & 52.837444 & -27.821638 & 0.9 & 1.113\tablenotemark{b}\phm{d}\phm{a} & \nodata\\ GALEX033255-281320\phn &        53.231551 &       -28.222466 & 22.46 & -27.50\phm{-} & 1.087 & 42.95 & 28$\pm$3 & \nodata & \nodata & 53.231567 & -28.222486 & 0.1 & 1.085\tablenotemark{b}\phm{d}\phm{a} & \nodata\\ GALEX033112-281442\tablenotemark{i} &        52.803095 &       -28.245108 & 22.19 & -28.20\phm{-} & 1.055 & 42.85 & 19$\pm$3 & \nodata & AGN & 52.801987 & -28.244835 & 3.6 & 1.042\tablenotemark{b}\phm{d}\phm{a} & \nodata\\ GALEX033413-275940\tablenotemark{i} &        53.555847 &       -27.994665 & 24.21 & 26.59 & 1.052 & 42.78 & 118$\pm$12 & \nodata & \nodata & 53.555477 & -27.994663 & 1.2 & 1.051\tablenotemark{b}\phm{d}\phm{a} & AGN\\ GALEX033212-274913\tablenotemark{i} &        53.051599 &       -27.820316 & 22.68 & 27.04 & 1.053 & 42.89 & 33$\pm$4 & \nodata & \nodata & 53.051613 & -27.820168 & 0.5 & 1.048\tablenotemark{b}\phm{d}\phm{a} & \nodata\\ GALEX033200-274319\phn &        53.001430 &       -27.722033 & 22.40 & 24.39 & 1.042 & 43.25 & 59$\pm$4 & -14.2 & \nodata & 53.001520 & -27.722071 & 0.3 & 1.041\tablenotemark{d}\phm{d}\phm{a} & AGN\\ GALEX033228-273614\phn &        53.121022 &       -27.604317 & 23.62 & 26.66 & 1.036 & 42.77 & 59$\pm$9 & \nodata & \nodata & 53.118469 & -27.604019 & 8.2 & 1.046\tablenotemark{b}\phm{d}\phm{a} & \nodata\\ GALEX033113-274949\phn &        52.807630 &       -27.830242 & 22.45 & 31.95 & 1.031 & 42.81 & 24$\pm$3 & \nodata & \nodata & 52.807545 & -27.830231 & 0.3 & 1.026\tablenotemark{b}\phm{d}\phm{a} & \nodata\\ GALEX033331-280625\tablenotemark{i} &        53.382523 &       -28.107164 & 24.41 & 27.66 & 1.029 & 42.80 & 141$\pm$14 & \nodata & \nodata & 53.382084 & -28.106623 & 2.4 & 1.026\tablenotemark{b}\phm{d}\phm{a} & AGN\\ GALEX033202-280320\phn &        53.010734 &       -28.055702 & 23.44 & 25.22 & 1.018 & 42.94 & 81$\pm$6 & -13.7 & AGN & 53.010658 & -28.055834 & 0.5 & 1.015\tablenotemark{b}\phm{d}\phm{a} & AGN\\ GALEX033111-275506\phn &        52.798577 &       -27.918436 & 23.81 & -27.52\phm{-} & 1.010 & 43.12 & 168$\pm$17 & \nodata & \nodata & 52.798473 & -27.918442 & 0.3 & 1.023\tablenotemark{b}\phm{d}\phm{a} & \nodata\\ GALEX033336-274224\phn &        53.402806 &       -27.706674 & 23.02 & 26.37 & 0.977 & 43.22 & 119$\pm$6 & -14.8 & AGN & 53.402771 & -27.706703 & 0.2 & 0.976\tablenotemark{b}\phm{d}\phm{a} & \nodata\\ GALEX033204-273725\phn &        53.017064 &       -27.623785 & 22.78 & 24.25 & 0.977 & 43.36 & 133$\pm$8 & -13.6 & AGN & 53.016788 & -27.623751 & 0.9 & 0.970\tablenotemark{e}\phm{d}\phm{a} & AGN\\ GALEX033246-274154\phn &        53.195680 &       -27.698408 & 23.17 & 26.61 & 0.976 & 43.20 & 128$\pm$10 & \nodata & \nodata & 53.196271 & -27.698547 & 2.1 & no z\phm{d}\phm{a} & \nodata\\ GALEX033342-274035\tablenotemark{i} &        53.425620 &       -27.676525 & 24.73 & 26.03 & 0.956 & 42.59 & 146$\pm$15 & \nodata & \nodata & 53.425163 & -27.676567 & 1.5 & 0.948\tablenotemark{b}\phm{d}\phm{a} & \nodata\\ GALEX033335-273934\phn &        53.398183 &       -27.659591 & 22.54 & 23.31 & 0.949 &   &   & -14.3 & \nodata & 53.398163 & -27.659760 & 0.6 & (0.622)\tablenotemark{b}\phm{d}\phm{a} & \nodata\\ GALEX033329-280127\phn &        53.373274 &       -28.024282 & 23.16 & 25.95 & 0.942 & 42.62 & 37$\pm$6 & \nodata & \nodata & 53.373272 & -28.024244 & 0.1 & 0.940\tablenotemark{b}\phm{d}\phm{a} & \nodata\\ GALEX033206-281408\phn &        53.026465 &       -28.235703 & 21.79 & 25.75 & 0.913 & 43.30 & 56$\pm$3 & \nodata & \nodata & 53.026493 & -28.235905 & 0.7 & 0.909\tablenotemark{b}\phm{d}\phm{a} & \nodata\\ GALEX033044-280237\phn &        52.686897 &       -28.043707 & 23.25 & 26.58 & 0.869 & 42.82 & 116$\pm$13 & \nodata & \nodata & 52.686813 & -28.043819 & 0.5 & 0.859\tablenotemark{b,c}\phm{d} & \nodata\\ GALEX033042-274215\tablenotemark{i} &        52.679011 &       -27.704257 & 23.03 & 25.18 & 0.861 & 42.57 & 38$\pm$4 & \nodata & \nodata & 52.678730 & -27.704462 & 1.2 & 0.856\tablenotemark{b}\phm{d}\phm{a} & \nodata\\ GALEX033409-280129\tablenotemark{i} &        53.541285 &       -28.024923 & 23.34 & 26.40 & 0.857 & 42.63 & 60$\pm$6 & \nodata & \nodata & 53.540737 & -28.024443 & 2.5 & 0.854\tablenotemark{b}\phm{d}\phm{a} & \nodata\\ GALEX033251-280809\phn &        53.214351 &       -28.135966 & 22.70 & 26.17 & 0.838 & 42.71 & 43$\pm$4 & \nodata & \nodata & 53.214489 & -28.136051 & 0.5 & 0.832\tablenotemark{b}\phm{d}\phm{a} & \nodata\\ GALEX033314-274834\phn &        53.310626 &       -27.809526 & 24.23 & 25.48 & 0.830 & 42.65 & 155$\pm$21 & \nodata & \nodata & 53.310975 & -27.809095 & 1.9 & 0.826\tablenotemark{d}\phm{d}\phm{a} & \nodata\\ GALEX033045-274506\phn &        52.689797 &       -27.751818 & 23.04 & 25.67 & 0.832 & 42.74 & 63$\pm$6 & \nodata & \nodata & 52.689950 & -27.752197 & 1.4 & no z\tablenotemark{h}\phm{a} & \nodata\\ GALEX033405-275023\tablenotemark{i} &        53.521292 &       -27.839993 & 23.13 & 25.03 & 0.825 & 42.56 & 47$\pm$9 & \nodata & \nodata & 53.520832 & -27.839972 & 1.5 & 0.821\tablenotemark{b}\phm{d}\phm{a} & \nodata\\ GALEX033057-273316\phn &        52.739776 &       -27.554542 & 22.37 & 24.09 & 0.794 & 42.73 & 38$\pm$3 & \nodata & \nodata & 52.739708 & -27.554583 & 0.3 & 0.790\tablenotemark{b}\phm{d}\phm{a} & \nodata\\ GALEX033124-275625\phn &        52.851698 &       -27.940441 & 22.69 & 25.82 & 0.778 & 42.57 & 36$\pm$4 & \nodata & \nodata & 52.851761 & -27.940475 & 0.2 & 0.773\tablenotemark{b}\phm{d}\phm{a} & \nodata\\ GALEX033235-274059\phn &        53.148932 &       -27.683184 & 23.67 & 25.70 & 0.745 & 42.79 & 182$\pm$15 & \nodata & \nodata & 53.149258 & -27.683273 & 1.1 & 0.735\tablenotemark{b}\phm{d}\phm{a} & \nodata\\ GALEX033146-274846\phn &        52.942503 &       -27.812779 & 22.98 & 25.88 & 0.742 & 42.75 & 86$\pm$6 & \nodata & \nodata & 52.942345 & -27.812771 & 0.5 & 0.736\tablenotemark{b}\phm{d}\phm{a} & \nodata\\ GALEX033100-273020\tablenotemark{i} &        52.751186 &       -27.505665 & 22.52 & 24.80 & 0.733 & 42.68 & 51$\pm$5 & \nodata & \nodata & 52.750877 & -27.505367 & 1.5 & 0.727\tablenotemark{b}\phm{d}\phm{a} & \nodata\\ GALEX033131-273429\phn &        52.880399 &       -27.574774 & 22.56 & 25.77 & 0.695 & 43.28 & 280$\pm$11 & -14.1 & \nodata & 52.880260 & -27.574841 & 0.5 & 0.688\tablenotemark{f}\phm{d}\phm{a} & AGN\\ GALEX033321-275031\tablenotemark{i} &        53.341285 &       -27.842052 & 23.16 & 25.06 & 0.691 & 42.59 & 90$\pm$9 & \nodata & AGN & \nodata & \nodata & \nodata & \nodata\phm{d}\phm{a} & \nodata\\ GALEX033150-274116\tablenotemark{i} &        52.962287 &       -27.687773 & 22.28 & 23.15 & 0.673 & 42.70 & 54$\pm$5 & -14.2 & AGN & 52.962346 & -27.687700 & 0.3 & 0.667\tablenotemark{d}\phm{d}\phm{a} & \nodata\\ \enddata \tablecomments{} \tablenotetext{a}{\rm{X-ray} data from \citet{lehmer05}} \tablenotetext{b}{\rm{This} paper's DEIMOS spectra} \tablenotetext{c}{\rm{This} paper's SALT spectra} \tablenotetext{d}{\rm{Redshifts} from \citet{cooper12}} \tablenotetext{e}{\rm{\citet{treister09}}; \citet{szokoly04} give $z=0.977$} \tablenotetext{f}{\rm{\citet{silverman10}}} \tablenotetext{g}{\rm{[alt] Alternative} counterpart at 52.942825, -27.494650, 2.2, (1.191)\tablenotemark{b} (R.A., Decl., Offset, $z_{\rm{opt}}$)} \tablenotetext{h}{\rm{[alt] Two} untargeted counterparts $\sim2''$ from NUV position} \tablenotetext{i}{\rm{LAE} candidate not listed in BCW12} \label{cdfs} \end{deluxetable} 
\clearpage
\begin{deluxetable}{ccccccccccccccc} \tabletypesize{\scriptsize}  \setlength{\tabcolsep}{0.02in} \tablecolumns{15} \tablewidth{0pc} \tablecaption{Emission-line Sample: GROTH-00} \tablehead{ \colhead{Name}    &  \colhead{R.A.} &   \colhead{Decl.} & \colhead{NUV} & \colhead{FUV} & \colhead{$z_{\rm{galex}}$} & \colhead{log $L$(Ly$\alpha$)} & \colhead{EW$_{\rm{r}}$(Ly$\alpha$)} & \colhead{log $f^{a}_{2-10~ \rm{keV}}$} & \colhead{Type(UV)} & \colhead{R.A.(opt)} &  \colhead{Decl.(opt)} & \colhead{Offset} & \colhead{$z_{\rm{opt}}$} & \colhead{Type(opt)}\\ \colhead{} & \colhead{(J2000.0)}   & \colhead{(J2000.0)} & \colhead{(AB)} & \colhead{(AB)} & \colhead{} & \colhead{(erg s$^{-1}$)} & \colhead{(\AA)} & \colhead{(erg cm$^{-2}$s$^{-1}$)} & \colhead{} & \colhead{(J2000.0)}   & \colhead{(J2000.0)} & \colhead{(arcsec)} & \colhead{} & \colhead{}\\ \colhead{(1)} & \colhead{(2)} & \colhead{(3)} & \colhead{(4)} & \colhead{(5)} & \colhead{(6)} & \colhead{(7)}  & \colhead{(8)}  & \colhead{(9)} & \colhead{(10)} & \colhead{(11)} & \colhead{(12)} & \colhead{(13)} & \colhead{(14)} & \colhead{(15)}} \startdata GALEX142028+524640 &        215.11946 &        52.777836 & 23.41 & 26.75 & 1.131 & 42.57 & 25$\pm$4 & \nodata & \nodata & 215.11989 & 52.777527 & 1.5 & 1.134\tablenotemark{b}\phm{d}\phm{a} & \nodata\\ GALEX141814+524415 &        214.56075 &        52.737531 & 23.97 & -27.83\phm{-} & 1.128 & 43.10 & 144$\pm$14 & \nodata & \nodata & 214.56084 & 52.737473 & 0.3 & 1.123\tablenotemark{b}\phm{d}\phm{a} & \nodata\\ GALEX141950+522542 &        214.96249 &        52.428532 & 24.09 & 26.34 & 1.114 & 42.97 & 124$\pm$12 & \nodata & \nodata & 214.96179 & 52.427998 & 2.5 & 1.111\tablenotemark{b}\phm{d}\phm{a} & \nodata\\ GALEX142013+524008 &        215.05510 &        52.669118 & 25.11 & 29.17 & 1.100 & 43.05 & 380$\pm$49 & \nodata & \nodata & 215.05558 & 52.669342 & 1.2 & no z\tablenotemark{d}\phm{a} & \nodata\\ GALEX141842+522919 &        214.67858 &        52.488791 & 22.69 & 23.68 & 1.084 & 42.98 & 37$\pm$5 & -14.3 & \nodata & 214.67863 & 52.488586 & 0.7 & 1.084\tablenotemark{b}\phm{d}\phm{a} & AGN\\ GALEX141824+522329 &        214.60416 &        52.391463 & 23.51 & 24.99 & 1.083 & 43.12 & 110$\pm$12 & -13.6 & \nodata & 214.60379 & 52.391666 & 1.1 & 1.081\tablenotemark{b}\phm{d}\phm{a} & AGN\\ GALEX141718+525521 &        214.32734 &        52.922402 & 22.47 & 24.41 & 1.067 & 43.58 & 127$\pm$4 & \nodata & AGN & 214.32666 & 52.922764 & 2.0 & 1.064\tablenotemark{b}\phm{d}\phm{a} & AGN\\ GALEX142028+522516 &        215.12077 &        52.421360 & 23.00 & 26.88 & 1.052 & 42.94 & 51$\pm$4 & \nodata & \nodata & 215.12045 & 52.421455 & 0.8 & 1.045\tablenotemark{b}\phm{d}\phm{a} & AGN\\ GALEX141842+523735 &        214.67549 &        52.626454 & 23.07 & 25.43 & 1.038 & 43.17 & 93$\pm$14 & \nodata & \nodata & 214.67545 & 52.625916 & 1.9 & 1.035\tablenotemark{b}\phm{d}\phm{a} & AGN\\ GALEX141737+524236 &        214.40603 &        52.709992 & 22.20 & 26.33 & 1.027 & 43.53 & 99$\pm$6 & -13.3 & \nodata & 214.40543 & 52.710182 & 1.5 & 1.025\tablenotemark{c}\phm{d}\phm{a} & AGN\\ GALEX141815+530452 &        214.56554 &        53.081148 & 22.36 & 23.26 & 1.028 & 43.07 & 41$\pm$3 & \nodata & \nodata & 214.56442 & 53.081223 & 2.4 & 1.020\tablenotemark{b}\phm{d}\phm{a} & AGN\\ GALEX142015+525024 &        215.06357 &        52.840222 & 24.04 & -28.66\phm{-} & 1.021 & 43.13 & 223$\pm$20 & \nodata & \nodata & 215.06401 & 52.840656 & 1.8 & 1.014\tablenotemark{b}\phm{d}\phm{a} & \nodata\\ GALEX142144+523627 &        215.43607 &        52.607699 & 21.83 & 24.66 & 1.011 & 43.32 & 46$\pm$3 & \nodata & \nodata & 215.43503 & 52.607677 & 2.3 & 1.006\tablenotemark{b}\phm{d}\phm{a} & AGN\\ GALEX142027+530455 &        215.11514 &        53.081989 & 23.18 & -27.68\phm{-} & 1.002 & 42.82 & 54$\pm$6 & -13.1 & \nodata & 215.11458 & 53.081806 & 1.4 & 0.994\tablenotemark{b}\phm{d}\phm{a} & AGN\\ GALEX141851+523600 &        214.71624 &        52.600051 & 23.54 & 26.00 & 1.000 & 43.00 & 112$\pm$14 & -14.1 & \nodata & 214.71530 & 52.599945 & 2.1 & 0.986\tablenotemark{b}\phm{d}\phm{a} & AGN\\ GALEX142242+525245 &        215.67896 &        52.879323 & 22.07 & -26.74\phm{-} & 1.003 & 42.65 & 12$\pm$3 & \nodata & \nodata & 215.67851 & 52.879360 & 1.0 & 0.999\tablenotemark{b}\phm{d}\phm{a} & \nodata\\ GALEX142033+530617 &        215.14121 &        53.104797 & 23.62 & 25.29 & 0.999 & 42.80 & 75$\pm$8 & -14.5 & \nodata & 215.14162 & 53.105114 & 1.5 & 0.994\tablenotemark{b}\phm{d}\phm{a} & \nodata\\ GALEX141902+525637 &        214.76161 &        52.943789 & 22.47 & 24.55 & 0.984 & 42.56 & 21$\pm$4 & \nodata & \nodata & 214.76091 & 52.943871 & 1.6 & 0.980\tablenotemark{b}\phm{d}\phm{a} & \nodata\\ GALEX142207+525411 &        215.52938 &        52.903111 & 23.07 & 25.35 & 0.996 & 43.16 & 104$\pm$5 & \nodata & \nodata & 215.52910 & 52.902931 & 0.9 & 0.994\tablenotemark{b}\phm{d}\phm{a} & AGN\\ GALEX142148+522902 &        215.45007 &        52.484033 & 23.04 & 27.40 & 0.961 & 43.23 & 131$\pm$13 & \nodata & \nodata & 215.44913 & 52.484470 & 2.6 & 0.961\tablenotemark{b}\phm{d}\phm{a} & \nodata\\ GALEX142244+525442 &        215.68588 &        52.911824 & 23.85 & 26.89 & 0.958 & 43.26 & 300$\pm$22 & \nodata & \nodata & 215.68564 & 52.911930 & 0.6 & 0.954\tablenotemark{b}\phm{d}\phm{a} & \nodata\\ GALEX141746+525259 &        214.44522 &        52.882997 & 24.17 & -27.28\phm{-} & 0.957 & 42.58 & 87$\pm$10 & \nodata & \nodata & 214.44400 & 52.884399 & 5.7 & 0.957\tablenotemark{b}\phm{d}\phm{a} & \nodata\\ GALEX141946+524755 &        214.94289 &        52.798752 & 22.23 & 28.91 & 0.928 & 42.21 & 6$\pm$3 & \nodata & \nodata & 214.94319 & 52.798855 & 0.7 & 0.914\tablenotemark{b}\phm{d}\phm{a} & \nodata\\ GALEX141935+524127 &        214.89632 &        52.690888 & 22.13 & 24.21 & 0.920 & 42.72 & 20$\pm$3 & \nodata & \nodata & 214.89662 & 52.690529 & 1.4 & 0.917\tablenotemark{b}\phm{d}\phm{a} & \nodata\\ GALEX141733+530403 &        214.39094 &        53.067620 & 23.21 & 24.87 & 0.918 & 43.07 & 120$\pm$10 & \nodata & \nodata & 214.39050 & 53.067913 & 1.4 & 0.915\tablenotemark{b}\phm{d}\phm{a} & AGN\\ GALEX141833+525525 &        214.64107 &        52.923739 & 23.46 & 26.32 & 0.916 & 42.74 & 72$\pm$7 & \nodata & \nodata & 214.64212 & 52.924011 & 2.5 & 0.912\tablenotemark{b}\phm{d}\phm{a} & \nodata\\ GALEX142201+530637 &        215.50453 &        53.110604 & 24.13 & 25.62 & 0.912 & 43.19 & 398$\pm$53 & -13.3 & \nodata & 215.50462 & 53.110195 & 1.5 & 0.898\tablenotemark{b}\phm{d}\phm{a} & AGN\\ GALEX142202+530823 &        215.50844 &        53.139988 & 22.85 & 24.75 & 0.872 &   &   & \nodata & \nodata & 215.50819 & 53.140114 & 1.3 & (0.572)\tablenotemark{b}\phm{d}\phm{a} & \nodata\\ GALEX142028+525839 &        215.11758 &        52.977690 & 23.30 & -27.77\phm{-} & 0.870 & 42.81 & 83$\pm$8 & -13.9 & \nodata & 215.11682 & 52.978155 & 2.3 & 0.871\tablenotemark{c}\phm{d}\phm{a} & \nodata\\ GALEX141800+522514 &        214.50369 &        52.420519 & 22.59 & 25.99 & 0.838 & 42.89 & 60$\pm$5 & \nodata & \nodata & 214.50280 & 52.419863 & 3.1 & 0.833\tablenotemark{c}\phm{d}\phm{a} & \nodata\\ GALEX141722+524209 &        214.34177 &        52.702458 & 22.26 & 26.40 & 0.838 & 43.07 & 65$\pm$5 & \nodata & \nodata & 214.34238 & 52.702278 & 1.5 & 0.838\tablenotemark{c}\phm{d}\phm{a} & \nodata\\ GALEX142202+522931 &        215.50918 &        52.492172 & 24.88 & 28.45 & 0.812 & 42.66 & 307$\pm$52 & \nodata & \nodata & 215.51004 & 52.493038 & 3.6 & 0.807\tablenotemark{b}\phm{d}\phm{a} & \nodata\\ GALEX142223+525642 &        215.59777 &        52.945235 & 23.55 & 26.86 & 0.770 & 42.90 & 187$\pm$21 & \nodata & \nodata & 215.59738 & 52.944958 & 1.3 & 0.763\tablenotemark{b}\phm{d}\phm{a} & \nodata\\ GALEX141858+523209 &        214.74578 &        52.535948 & 21.76 & 22.67 & 0.762 & 43.11 & 57$\pm$4 & -14.4 & \nodata & 214.74509 & 52.536110 & 1.6 & 0.758\tablenotemark{b}\phm{d}\phm{a} & AGN\\ GALEX142049+523203 &        215.20764 &        52.534438 & 23.62 & 28.61 & 0.754 & 43.03 & 273$\pm$21 & \nodata & \nodata & 215.20830 & 52.534378 & 1.5 & 0.758\tablenotemark{b}\phm{d}\phm{a} & \nodata\\ GALEX142212+522634 &        215.55384 &        52.443058 & 23.67 & 27.45 & 0.762 & 42.47 & 79$\pm$11 & \nodata & \nodata & 215.55397 & 52.443161 & 0.5 & 0.759\tablenotemark{b}\phm{d}\phm{a} & \nodata\\ GALEX142057+525642 &        215.24117 &        52.945171 & 22.54 & 24.98 & 0.756 & 43.00 & 101$\pm$6 & \nodata & \nodata & 215.24106 & 52.944958 & 0.8 & 0.747\tablenotemark{b}\phm{d}\phm{a} & \nodata\\ GALEX141746+525646 &        214.44476 &        52.946049 & 23.28 & 25.21 & 0.750 & 42.97 & 191$\pm$22 & \nodata & \nodata & 214.44504 & 52.946918 & 3.2 & 0.746\tablenotemark{b}\phm{d}\phm{a} & AGN\\ GALEX141842+522256 &        214.67524 &        52.382180 & 23.65 & 26.29 & 0.750 & 42.49 & 85$\pm$16 & \nodata & \nodata & 214.67545 & 52.381447 & 2.7 & 0.747\tablenotemark{b}\phm{d}\phm{a} & \nodata\\ GALEX141842+530140 &        214.67697 &        53.027806 & 23.03 & 25.01 & 0.748 & 42.59 & 62$\pm$9 & \nodata & \nodata & 214.67871 & 53.028191 & 4.0 & 0.741\tablenotemark{b}\phm{d}\phm{a} & \nodata\\ GALEX142139+523401 &        215.41400 &        52.567205 & 23.82 & 26.24 & 0.741 &   &   & \nodata & AGN & 215.41359 & 52.568192 & 3.8 & (0.692)\tablenotemark{b}\phm{d}\phm{a} & \nodata\\ \enddata \tablecomments{} \tablenotetext{a}{\rm{X-ray} data from \citet{laird09}} \tablenotetext{b}{\rm{This} paper's DEIMOS spectra} \tablenotetext{c}{\rm{Archival DEIMOS} DEEP2 spectra \citep{newman12}} \tablenotetext{d}{\rm{[alt] Alternative} counterpart at 215.05445, 52.667969, 4.4, no z (R.A., Decl., Offset, $z_{\rm{opt}}$)} \label{groth} \end{deluxetable} 
\clearpage
\begin{deluxetable}{ccccccccccccccc} \tabletypesize{\scriptsize}  \setlength{\tabcolsep}{0.02in} \tablecolumns{15} \tablewidth{0pc} \tablecaption{Emission-line Sample: NGPDWS-00} \tablehead{ \colhead{Name}    &  \colhead{R.A.} &   \colhead{Decl.} & \colhead{NUV} & \colhead{FUV} & \colhead{$z_{\rm{galex}}$} & \colhead{log $L$(Ly$\alpha$)} & \colhead{EW$_{\rm{r}}$(Ly$\alpha$)} & \colhead{log $f^{a}_{2-7~  \rm{keV}}$} & \colhead{Type(UV)} &  \colhead{R.A.(opt)} &  \colhead{Decl.(opt)} & \colhead{Offset} & \colhead{$z_{\rm{opt}}$} & \colhead{Type(opt)}\\ \colhead{} & \colhead{(J2000.0)}   & \colhead{(J2000.0)} & \colhead{(AB)} & \colhead{(AB)} & \colhead{} & \colhead{(erg s$^{-1}$)} & \colhead{(\AA)} & \colhead{(erg cm$^{-2}$s$^{-1}$)} & \colhead{} & \colhead{(J2000.0)}   & \colhead{(J2000.0)} & \colhead{(arcsec)} & \colhead{} & \colhead{}\\ \colhead{(1)} & \colhead{(2)} & \colhead{(3)} & \colhead{(4)} & \colhead{(5)} & \colhead{(6)} & \colhead{(7)}  & \colhead{(8)}  & \colhead{(9)} & \colhead{(10)} & \colhead{(11)} & \colhead{(12)} & \colhead{(13)} & \colhead{(14)} & \colhead{(15)}} \startdata GALEX143557+350702 &        218.98764 &        35.117445 & 23.82 & -28.46\phm{-} & 1.158 & 43.27 & 169$\pm$26 & \nodata\phm{f} & \nodata & 218.98735 & 35.116795 & 2.5 & 1.152\tablenotemark{b}\phm{d} & \nodata\\ GALEX143539+351001 &        218.91250 &        35.167233 & 22.20 & 25.99 & 1.083 & 43.40 & 63$\pm$6 & \nodata\phm{f} & \nodata & 218.91272 & 35.167511 & 1.2 & 1.082\tablenotemark{b}\phm{d} & \nodata\\ GALEX143545+350039 &        218.93744 &        35.011330 & 24.17 & 26.63 & 1.063 &   &   & \nodata\phm{f} & AGN & 218.93967 & 35.010777 & 5.9 & (0.722)\tablenotemark{b}\phm{d} & \nodata\\ GALEX143520+350414 &        218.83421 &        35.070876 & 21.55 & 22.69 & 1.055 & 43.32 & 31$\pm$3 & -14.5\phm{f} & \nodata & 218.83405 & 35.070351 & 2.0 & 1.052\tablenotemark{b}\phm{d} & AGN\\ GALEX143824+352325 &        219.60174 &        35.390787 & 21.92 & 24.92 & 1.039 & 42.92 & 18$\pm$3 & \nodata\phm{f} & AGN & 219.60257 & 35.390755 & 2.4 & 1.031\tablenotemark{b}\phm{d} & \nodata\\ GALEX143512+352338 &        218.80005 &        35.394162 & 24.84 & -31.03\phm{-} & 1.037 &   &   & \nodata\phm{f} & \nodata & 218.80109 & 35.392956 & 4.2 & (1.080)\tablenotemark{b}\tablenotemark{,c} & \nodata\\ GALEX143726+351448 &        219.36103 &        35.247140 & 22.44 & 24.39 & 1.034 & 43.13 & 47$\pm$6 & -15.1\tablenotemark{e} & \nodata & \nodata & \nodata & \nodata & \nodata\phm{d} & \nodata\\ GALEX143443+353037 &        218.68009 &        35.510756 & 23.39 & 28.50 & 1.034 & 42.85 & 61$\pm$8 & \nodata\phm{f} & \nodata & 218.68036 & 35.510929 & 1.0 & 1.025\tablenotemark{b}\phm{d} & \nodata\\ GALEX143750+350645 &        219.45954 &        35.112962 & 22.22 & 23.70 & 1.010 &   &   & -15.0\phm{f} & \nodata & 219.46024 & 35.113098 & 1.9 & (0.577)\tablenotemark{b}\phm{d} & \nodata\\ GALEX143449+350248 &        218.70428 &        35.046988 & 23.21 & 29.53 & 1.009 & 43.60 & 303$\pm$9 & -15.3\phm{f} & \nodata & \nodata & \nodata & \nodata & \nodata\phm{d} & \nodata\\ GALEX143810+350424 &        219.54441 &        35.073741 & 23.73 & 26.81 & 1.006 & 43.15 & 177$\pm$14 & \nodata\phm{f} & \nodata & 219.54455 & 35.073328 & 0.7 & no z\phm{d} & \nodata\\ GALEX143450+352520 &        218.70844 &        35.422650 & 22.85 & 26.74 & 0.967 & 43.29 & 122$\pm$7 & -14.2\phm{f} & \nodata & 218.70876 & 35.422508 & 1.1 & 0.964\tablenotemark{b}\phm{d} & AGN\\ GALEX143512+345907 &        218.79986 &        34.985750 & 24.58 & 28.79 & 0.963 &   &   & \nodata\phm{f} & \nodata & 218.80116 & 34.984894 & 3.8 & (0.884)\tablenotemark{b}\phm{d} & \nodata\\ GALEX143510+351602 &        218.79406 &        35.267619 & 23.42 & 26.28 & 0.941 &   &   & \nodata\phm{f} & \nodata & 218.79521 & 35.267330 & 2.5 & (0.478)\tablenotemark{b}\tablenotemark{,d} & \nodata\\ GALEX143639+350557 &        219.16252 &        35.099672 & 22.31 & 23.43 & 0.937 & 42.87 & 32$\pm$3 & \nodata\phm{f} & \nodata & 219.16257 & 35.099346 & 1.2 & 0.932\tablenotemark{b}\phm{d} & \nodata\\ GALEX143702+350155 &        219.26169 &        35.032282 & 22.94 & 24.99 & 0.933 & 43.34 & 149$\pm$10 & -14.6\phm{f} & \nodata & 219.26167 & 35.032078 & 0.7 & 0.927\tablenotemark{b}\phm{d} & AGN\\ GALEX143531+351415 &        218.88092 &        35.237743 & 23.09 & 27.50 & 0.934 & 42.96 & 78$\pm$11 & \nodata\phm{f} & \nodata & 218.88094 & 35.238354 & 2.2 & 0.930\tablenotemark{b}\phm{d} & \nodata\\ GALEX143730+352016 &        219.37524 &        35.338103 & 22.51 & 24.34 & 0.907 & 42.90 & 46$\pm$5 & \nodata\phm{f} & \nodata & 219.37563 & 35.337757 & 1.7 & 0.904\tablenotemark{b}\phm{d} & AGN\\ GALEX143521+350509 &        218.83940 &        35.086214 & 21.89 & 26.57 & 0.901 &   &   & \nodata\phm{f} & \nodata & 218.83961 & 35.085414 & 2.1 & star\phm{d} & star\\ GALEX143819+351542 &        219.58154 &        35.262070 & 23.01 & 26.70 & 0.889 & 42.63 & 39$\pm$9 & \nodata\phm{f} & \nodata & 219.58224 & 35.261349 & 3.3 & 0.889\tablenotemark{b}\phm{d} & \nodata\\ GALEX143652+350537 &        219.21699 &        35.093944 & 21.90 & 23.03 & 0.870 & 43.25 & 63$\pm$5 & -15.0\phm{f} & \nodata & 219.21677 & 35.093887 & 0.7 & 0.866\tablenotemark{b}\phm{d} & AGN\\ GALEX143729+350732 &        219.37419 &        35.125830 & 22.68 & 24.16 & 0.868 & 43.31 & 151$\pm$11 & -15.0\phm{f} & \nodata & 219.37398 & 35.126053 & 1.0 & 0.860\tablenotemark{b}\phm{d} & AGN\\ GALEX143457+353213 &        218.73762 &        35.537287 & 21.75 & 23.02 & 0.865 & 42.82 & 21$\pm$3 & -14.9\phm{f} & \nodata & 218.73782 & 35.537319 & 0.6 & 0.857\tablenotemark{b}\phm{d} & AGN\\ GALEX143715+353355 &        219.31319 &        35.565674 & 22.38 & 23.84 & 0.854 & 43.27 & 127$\pm$11 & -15.1\tablenotemark{e} & \nodata & 219.31288 & 35.565460 & 1.2 & 0.846\tablenotemark{b}\phm{d} & AGN\\ GALEX143459+352504 &        218.74643 &        35.418085 & 22.48 & -28.10\phm{-} & 0.848 &   &   & \nodata\phm{f} & \nodata & 218.74582 & 35.417044 & 4.0 & star\phm{d} & star\\ GALEX143800+352206 &        219.50182 &        35.368824 & 22.24 & 24.81 & 0.829 & 42.81 & 35$\pm$5 & -15.4\phm{f} & \nodata & \nodata & \nodata & \nodata & \nodata\phm{d} & \nodata\\ GALEX143642+345027 &        219.17868 &        34.841281 & 24.09 & 27.73 & 0.823 & 42.91 & 254$\pm$49 & \nodata\phm{f} & \nodata & 219.17896 & 34.841167 & 0.9 & 0.818\tablenotemark{b}\phm{d} & \nodata\\ GALEX143655+344610 &        219.23087 &        34.769857 & 24.20 & 28.61 & 0.820 & 42.88 & 269$\pm$30 & \nodata\phm{f} & \nodata & 219.23146 & 34.769501 & 2.2 & 0.813\tablenotemark{b}\phm{d} & \nodata\\ GALEX143801+352533 &        219.50452 &        35.426310 & 21.62 & 22.75 & 0.806 & 43.09 & 42$\pm$3 & -14.7\phm{f} & \nodata & \nodata & \nodata & \nodata & \nodata\phm{d} & \nodata\\ GALEX143700+353132 &        219.25323 &        35.525830 & 22.50 & 24.40 & 0.791 & 42.96 & 75$\pm$7 & \nodata\phm{f} & \nodata & 219.25333 & 35.525555 & 1.0 & 0.784\tablenotemark{b}\phm{d} & AGN\\ GALEX143515+353203 &        218.81441 &        35.534506 & 21.77 & 23.47 & 0.789 &   &   & \nodata\phm{f} & \nodata & 218.81459 & 35.534233 & 0.5 & (0.410)\tablenotemark{b}\phm{d} & \nodata\\ GALEX143821+350707 &        219.59055 &        35.118996 & 21.62 & 23.15 & 0.741 & 43.23 & 75$\pm$5 & -14.8\phm{f} & AGN & 219.59068 & 35.119396 & 1.5 & 0.728\tablenotemark{b}\phm{d} & \nodata\\ GALEX143829+352308 &        219.62144 &        35.385905 & 22.19 & 24.26 & 0.720 & 42.49 & 25$\pm$4 & \nodata\phm{f} & \nodata & 219.62173 & 35.385715 & 1.1 & 0.710\tablenotemark{b}\phm{d} & \nodata\\ GALEX143716+352324 &        219.32011 &        35.390438 & 24.61 & 27.67 & 0.716 & 42.61 & 365$\pm$65 & \nodata\phm{f} & \nodata & 219.31993 & 35.391083 & 2.4 & 0.699\tablenotemark{b}\phm{d} & \nodata\\ GALEX143813+351121 &        219.55694 &        35.189666 & 24.20 & 27.14 & 0.710 & 42.67 & 199$\pm$53 & \nodata\phm{f} & \nodata & 219.55760 & 35.190166 & 2.6 & 0.717\tablenotemark{b}\phm{d} & \nodata\\ \enddata \tablecomments{} \tablenotetext{a}{\rm{X-ray} data from \citet{kenter05}} \tablenotetext{b}{\rm{This} paper's DEIMOS spectra} \tablenotetext{c}{\rm{[alt] Alternative} counterpart at 218.98737, 35.115852, 5.8, (1.096)\tablenotemark{b} (R.A., Decl., Offset, $z_{\rm{opt}}$)} \tablenotetext{d}{\rm{[alt] Alternative} counterpart at 218.79877, 35.394573, 4.0, (0.977)\tablenotemark{b} (R.A., Decl., Offset, $z_{\rm{opt}}$)} \tablenotetext{e}{\rm{Source} not detected in X-ray hard band but detected in the soft 0.5-2 keV flux band} \label{ngpdws} \end{deluxetable} 
\clearpage
\begin{deluxetable}{ccccccccccccccc} \tabletypesize{\scriptsize}  \setlength{\tabcolsep}{0.02in} \tablecolumns{15} \tablewidth{0pc} \tablecaption{Emission-line Sample: COSMOS-00} \tablehead{ \colhead{Name}    &  \colhead{R.A.} &   \colhead{Decl.} & \colhead{NUV} & \colhead{FUV} & \colhead{$z_{\rm{galex}}$} & \colhead{log $L$(Ly$\alpha$)} & \colhead{EW$_{\rm{r}}$(Ly$\alpha$)} & \colhead{log $f^{a}_{2-10~  \rm{keV}}$} & \colhead{Type(UV)} & \colhead{R.A.(opt)} &  \colhead{Decl.(opt)} & \colhead{Offset} & \colhead{$z_{\rm{opt}}$} & \colhead{Type(opt)}\\ \colhead{} & \colhead{(J2000.0)}   & \colhead{(J2000.0)} & \colhead{(AB)} & \colhead{(AB)} & \colhead{} & \colhead{(erg s$^{-1}$)} & \colhead{(\AA)} & \colhead{(erg cm$^{-2}$s$^{-1}$)} & \colhead{} & \colhead{(J2000.0)}   & \colhead{(J2000.0)} & \colhead{(arcsec)} & \colhead{} & \colhead{}\\ \colhead{(1)} & \colhead{(2)} & \colhead{(3)} & \colhead{(4)} & \colhead{(5)} & \colhead{(6)} & \colhead{(7)}  & \colhead{(8)}  & \colhead{(9)} & \colhead{(10)} & \colhead{(11)} & \colhead{(12)} & \colhead{(13)} & \colhead{(14)} & \colhead{(15)}} \startdata GALEX095954+021707 &        149.97779 &        2.2852787 & 22.17 & 23.93 & 1.163 & 43.24 & 30$\pm$4 & -14.7\phm{f} & \nodata & \nodata & \nodata & \nodata & \nodata\phm{a} & \nodata\\ GALEX100145+023237 &        150.43904 &        2.5436927 & 23.59 & 24.84 & 1.154 & 43.35 & 158$\pm$13 & \nodata\phm{f} & \nodata & 150.43942 & 2.5435109 & 1.5 & 1.149\tablenotemark{b}\phm{a} & AGN\\ GALEX100016+015104 &        150.06776 &        1.8512265 & 21.74 & 23.40 & 1.140 & 43.63 & 57$\pm$4 & -13.6\phm{f} & AGN & 150.06784 & 1.8513390 & 0.5 & 1.134\tablenotemark{b}\phm{a} & AGN\\ GALEX100031+023012 &        150.13260 &        2.5034525 & 22.72 & 25.38 & 1.103 & 43.21 & 59$\pm$6 & \nodata\phm{f} & \nodata & 150.13235 & 2.5033109 & 1.0 & 1.099\tablenotemark{b,c} & \nodata\\ GALEX095918+014933 &        149.82753 &        1.8260336 & 22.89 & 25.11 & 1.070 & 43.38 & 109$\pm$10 & \nodata\phm{f} & \nodata & 149.82774 & 1.8258640 & 1.0 & 1.069\tablenotemark{b,c} & \nodata\\ GALEX095916+015048 &        149.81704 &        1.8467845 & 21.94 & 24.30 & 1.034 & 43.20 & 33$\pm$4 & -14.0\phm{f} & AGN & 149.81688 & 1.8467280 & 0.6 & 1.034\tablenotemark{b}\phm{a} & AGN\\ GALEX100141+021029 &        150.42197 &        2.1748164 & 21.83 & 23.00 & 0.989 & 43.47 & 65$\pm$5 & -13.5\phm{f} & AGN & 150.42221 & 2.1754150 & 2.3 & 0.982\tablenotemark{d}\phm{a} & AGN\\ GALEX100150+020936 &        150.45907 &        2.1598448 & 22.75 & 25.61 & 0.927 & 43.11 & 79$\pm$7 & \nodata\phm{f} & \nodata & 150.45946 & 2.1603360 & 2.3 & 0.926\tablenotemark{b,c} & AGN\\ GALEX100124+021447 &        150.34970 &        2.2463910 & 22.34 & 23.46 & 0.900 & 43.23 & 77$\pm$8 & -13.8\phm{f} & \nodata & 150.34988 & 2.2461319 & 1.1 & 0.894\tablenotemark{d}\phm{a} & \nodata\\ GALEX100202+020145 &        150.51078 &        2.0290676 & 22.95 & 24.72 & 0.894 & 43.20 & 124$\pm$9 & -13.6\phm{f} & \nodata & 150.51064 & 2.0292540 & 0.8 & 0.898\tablenotemark{d}\phm{a} & AGN\\ GALEX100049+021707 &        150.20630 &        2.2852412 & 23.88 & 24.96 & 0.877 & 42.96 & 181$\pm$30 & -13.7\phm{f} & \nodata & 150.20627 & 2.2857921 & 2.0 & 0.874\tablenotemark{d}\phm{a} & AGN\\ GALEX100136+020653 &        150.40266 &        2.1148013 & 22.28 & 25.06 & 0.853 & 42.92 & 42$\pm$5 & \nodata\phm{f} & \nodata & 150.40346 & 2.1155031 & 3.8 & 0.849\tablenotemark{b,c} & \nodata\\ GALEX100133+015451 &        150.39033 &        1.9141639 & 21.85 & 24.96 & 0.848 & 43.02 & 37$\pm$3 & \nodata\phm{f} & \nodata & 150.39020 & 1.9144530 & 1.1 & 0.844\tablenotemark{b,c} & \nodata\\ GALEX100002+021628 &        150.00988 &        2.2746592 & 21.90 & 23.78 & 0.847 & 42.99 & 34$\pm$4 & -13.5\phm{f} & \nodata & 150.00922 & 2.2755051 & 3.9 & 0.850\tablenotemark{d}\phm{a} & AGN\\ GALEX100207+021119 &        150.53190 &        2.1886307 & 22.31 & 22.60 & 0.838 & 43.25 & 101$\pm$19 & -14.0\phm{f} & \nodata & 150.53186 & 2.1889460 & 1.1 & 0.830\tablenotemark{d}\phm{a} & \nodata\\ GALEX100113+022548 &        150.30739 &        2.4301342 & 22.76 & 23.40 & 0.757 &   &   & -13.6\phm{f} & \nodata & 150.30808 & 2.4300640 & 2.5 & (0.374)\tablenotemark{e}\phm{a} & AGN\\ GALEX100029+022129 &        150.12442 &        2.3581919 & 22.26 & 23.42 & 0.739 & 42.93 & 66$\pm$9 & -13.6\phm{f} & AGN & 150.12370 & 2.3582580 & 2.6 & 0.728\tablenotemark{e}\phm{a} & AGN\\ GALEX095910+020732 &        149.79323 &        2.1258027 & 22.12 & 22.22 & 0.730 &   &   & -14.8\tablenotemark{f} & \nodata & 149.79295 & 2.1256490 & 1.1 & (0.353)\tablenotemark{d}\phm{a} & \nodata\\ GALEX100017+020013 &        150.07274 &        2.0037061 & 22.69 & 23.04 & 0.729 &   &   & -13.6\phm{f} & \nodata & 150.07301 & 2.0035551 & 1.1 & (0.350)\tablenotemark{d}\phm{a} & AGN\\ 
  \enddata \tablecomments{} \tablenotetext{a}{\rm{X-ray} data from \citet{elvis09}} \tablenotetext{b}{\rm{This} paper's DEIMOS spectra} \tablenotetext{c}{\rm{This} paper's SALT spectra} \tablenotetext{d}{\rm{Archival} Magellan spectra from \citet{trump09}} \tablenotetext{e}{\rm{Archival} VLT spectra from \citet{lilly07}} \tablenotetext{f}{\rm{Source} not detected in X-ray hard band but detected in the soft 0.5-2 keV flux band} \label{cosmos} \end{deluxetable} 

  \clearpage \end{landscape}

\end{document}